\documentclass{llncs}
\usepackage{cite}
\usepackage{amsmath,amssymb,amsfonts}
\usepackage{algorithmic}
\usepackage{caption}
\usepackage{graphicx}
\usepackage{textcomp}
\usepackage{multirow}
\usepackage{booktabs}
\usepackage{subcaption}

\usepackage[hyphens]{url}
\usepackage{comment}
\usepackage{float}

\newcommand{\CS}{Citizen Science} 
\newcommand{\pdf}{PDF}
\newcommand{\zip}{ZIP}
\newcommand{\csv}{CSV}

\def\BibTeX{{\rm B\kern-.05em{\sc i\kern-.025em b}\kern-.08em
    T\kern-.1667em\lower.7ex\hbox{E}\kern-.125emX}}
\begin{document}

\title{Analyzing social media with crowdsourcing in Crowd4SDG 
\thanks{This work was supported European Commission H2020 project  Crowd4SDG ``Citizen Science for Monitoring Climate Impacts and Achieving Climate Resilience'', No. 872944. Corresponding author: Barbara Pernici, ORCID: 0000-0002-2034-9774.}}

\author{
Carlo Bono\inst{1}\and
Mehmet O\u{g}uz  M\"{u}l\^{a}y\.{i}m\inst{2}\and
Cinzia Cappiello\inst{1}\and
Mark Carman\inst{1}\and
Jesus Cerquides\inst{2}\and 
Jose Luis Fernandez-Marquez\inst{3}\and
Rosy Mondardini\inst{4}\and
Edoardo Ramalli\inst{1}\and
Barbara Pernici*\inst{1}
}
\institute{
Politecnico di Milano, DEIB, Piazza Leonardo da Vinci 32, 20133 Milano, Italy \\
\email{[name.lastname]@polimi.it}
\and
Artificial Intelligence Research Institute (IIIA-CSIC)\\
\email{[oguz,cerquide]@iiia.csic.es}
\and
University of Geneva\\
\email{joseluis.fernandez@unige.ch}
\and
Citizen Science Center Zurich (UZH and ETHZ) \\
\email{maria.mondardini@uzh.ch}
}

\authorrunning{C. Bono et al.}
\titlerunning{Analyzing social media with crowdsourcing in Crowd4SDG}

\maketitle

\begin{abstract}
Social media have the potential to provide timely information about emergency situations and sudden events. However, finding relevant information among  millions of posts being posted every day can be difficult, and developing a data analysis project usually requires time and technical skills.

This study presents an approach that provides  flexible support for analyzing social media,  particularly during emergencies.  Different use cases in which social media analysis can be adopted are introduced, and the challenges of retrieving information from large sets of posts are discussed. 

The focus is on analyzing images and text contained in social media posts and  a set of automatic data processing tools for filtering, classification, and geolocation of content with a human-in-the-loop approach to support the data analyst.
Such support includes both feedback and suggestions to configure automated tools, and crowdsourcing to gather inputs from citizens.
The results are validated by discussing three case studies developed within the Crowd4SDG H2020 European project.\\
%\ \\
%XXX To check the paper at the end \url{https://preflight.paperpal.com/partner/ieee/access}

%Abstract 100-250 words
\end{abstract}


\section{Introduction} \label{sec:intro} 

The extraction of information from digital sources is challenging. In particular, when extracting information from
social media, it is necessary to analyze large volumes of data with a wide variety of information, made of short and informal text with no structure and possibly containing noise, that is, information that is not useful or confusing. On the other hand, social media often document ongoing facts and can provide very timely information on situations and events, as well as images that can support knowledge extraction.

Citizens play an important role in providing information \CS~(CS). Several ways of contributing have been studied \cite{fritz2019citizen}. In particular, the
role of citizens can range from actively providing ideas and evidence to decision makers, to supporting data analysis tasks, and also to contributing indirectly, posting information on social media platforms from where it can be retrieved through information discovery techniques.

In the Crowd4SDG (Citizen Science for Monitoring Climate Impacts and Achieving Climate Resilience) H2020 European project, we studied an innovative approach to support Citizen Science data managers and analysts in social media analysis. 
The approach advocates the active role of citizen scientists in all phases of social media analysis. Specifically, we propose a data analysis pipeline in which AI-based data analysis tools are combined with crowdsourcing.
CS project managers and data analysts are in control of all phases, selecting the information sources and how to crawl them, providing input in the configuration of %the
automatic analysis tools based on the results, and allowing citizens to contribute  ideas and annotations. The final goal is to help tackle critical issues related to Sustainable Development Goals (SDG) by mining the collaboration and contributions of citizen scientists.
In particular, in Crowd4SDG, we focused on SDG~13 Climate action and on indicators to reduce the impact of emergency events on the population.
The project studied different modalities to involve citizens and developed an approach to collect and refine ideas, proposing challenges on specific themes for citizens\footnote{Some Crowd4SDG challenges are presented in \url{https://crowd4sdg.eu/gear-cycle-01/} and  \url{https://crowd4sdg.eu/gear-cycle-02/}}. A set of  challenges was proposed each year of the project, focusing on the combination of SDG~13 with a second SDG (SDG~11, SDG~5, and SDG~16 respectively), inviting citizen scientists to propose project ideas and helping refine them through an idea development program in which both project proposals were refined and tools for their realization were evaluated. 
In general, crowdsourcing poses several difficulties related to engaging and retaining participants \cite{wald2016design} which are not different when used to support social media analysis. In this case, some of the main problems are finding crowd workers willing to contribute, in particular for unpaid contributions, and giving them the right tasks to perform. 
Thus, some of the research efforts focused on the need to be effective in assigning tasks to citizens and were directed towards providing only relevant posts for analysis.
%, which was as well achieved by leveraging crowdsourcing. 
Another focus of research was being able to assess the consensus of crowd workers on assigned tasks, for which methods for advanced consensus analysis were proposed.

The result of our efforts to address the challenges in crowdsourcing is a set of dedicated tools called the ``Citizen Science Solution Kit'' (CSSK) including VisualCit for social media data filtering, the CS Project Builder (CSPB) for crowdsourcing social  media data analysis, and crowd analysis (Crowdnalysis).
The aim of this paper is to present the main results of Crowd4SDG %in developing tools 
to support CS project managers in developing \CS projects and to provide them with a flexible platform that allows for maintaining full control over the project without the need for technical interventions during its realization, and to provide flexible ways to combine different tools for a variety of analysis goals.
This paper discusses how the components of the CSSK can be used and combined   flexibly  and how they can be configured for the needs of specific projects in an easy way directly by CS Project Managers. 

The remainder of this paper is organized as follows. In Section \ref{sec:scenario}, we describe the general approach to including \CS in the development of social media analysis pipelines. Section \ref{sec:cssk} presents the Crowd4SDG CSSK, including the presentation of the types of tools proposed, a methodological approach for using them, the architecture of tool composition, and a detailed illustration of each tool. Section \ref{sec:validation} describes a validation approach for CS projects realized with the CSSK and illustrates and compares three different case studies. In Section \ref{sec:sota}, the state of the art is discussed and finally conclusions and future research directions are discussed in Section \ref{sec:conclusions}.

\section{Motivating scenario} \label{sec:scenario}

Social media have been investigated in many crisis and emergency scenarios, for two main reasons: i) they provide a viable source of information covering areas and situations in which it is difficult to set up systematic data collection (i.e.,  they are one of the possible ``non-traditional  data sources'' \cite{fritz2019citizen}); and ii) there is the potential to extract timely information concerning the affected areas.
Consequently, many research papers are published in this domain, including a special track at the International Conference on Information Systems for Crisis Response and Management on Social Media for Crisis Management \cite{special-track}.
One of the main goals of the research in this area is to automatically extract information about an event to characterize, highlight and react to emergencies. This domain particularly seeks visual information  because it provides substantial evidence of ongoing events \cite{havas2017e2mc, asif2021automatic}.

\begin{figure}[h]
    \centering
    \includegraphics[width=0.99\linewidth]{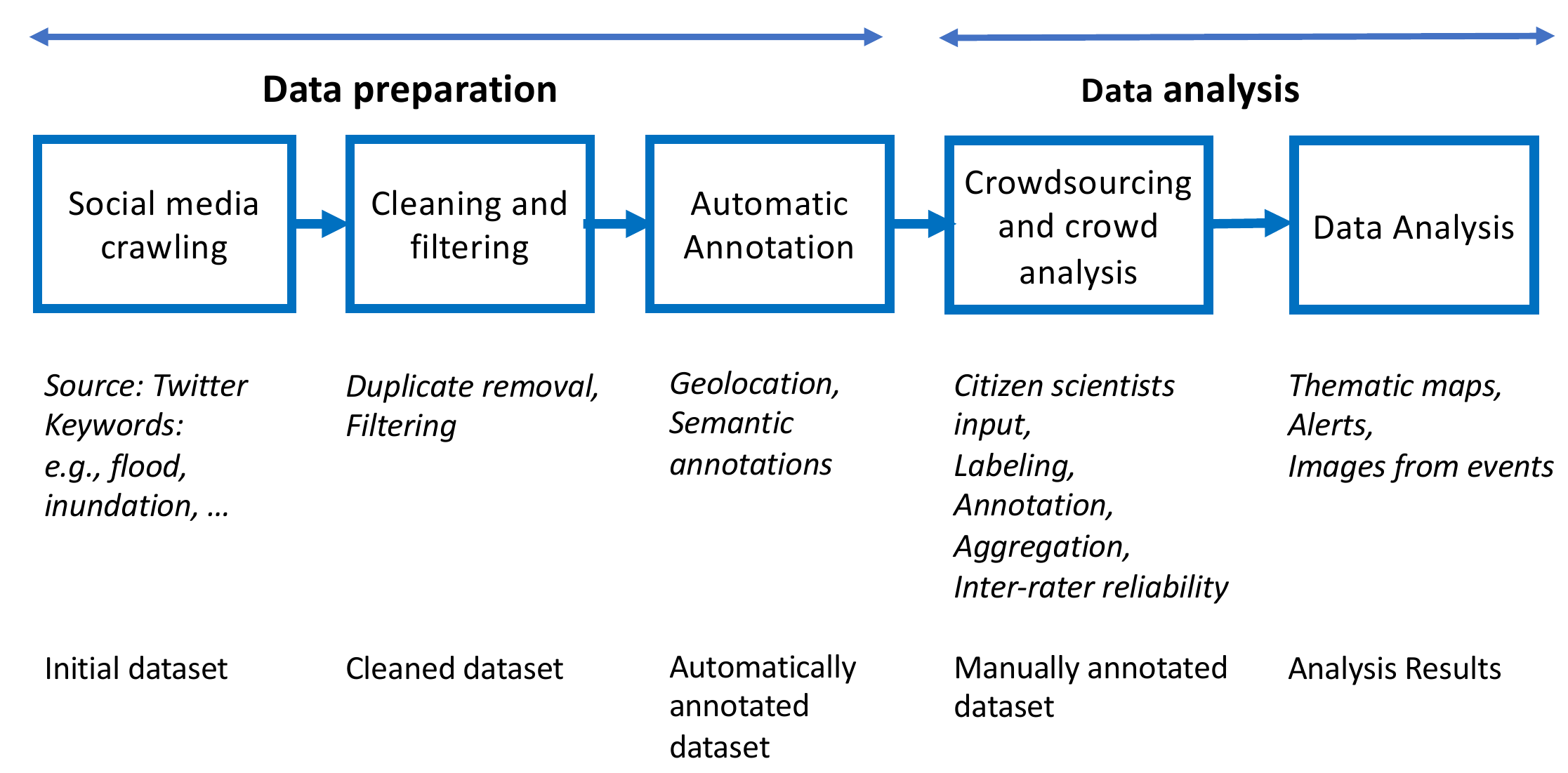}
    \caption{Social media analysis scenario.}
    \label{fig:pipeline-intro}
\end{figure}

Extracting relevant information from tweets presents several challenges. First of all, the contents of the tweets must be relevant to the event and to the type of information being searched.
Relevance can depend on the use of the information and the actors involved; therefore,  its definition is part of a data analysis project description, as it will be discussed in the following of this paper.
The extraction process requires several phases of data preparation and analysis, as raw datasets extracted from social media usually contain a very low percentage of useful information, estimated to be between 0.5\% and 3\% in the literature \cite{DBLP:conf/icse/NegriSARSSFCP21, scalia2021cime}. However, even a very small number of relevant posts can be useful in providing a significant overview on the areas in which the event occurs \cite{fohringer2015social,scotti2020enhanced}. To solve this ``needle in the haystack'' problem, a purely manual analysis is not viable, given the massive number of posts to be examined.
In Fig. \ref{fig:pipeline-intro}, a typical social media analysis scenario is depicted, in which a human computing approach is used, interleaving fully automatic and human-supported phases. The information extraction process can be split into two macrophases: in the first one, the \emph{Data preparation} phase, data are prepared for the analysis, while in the second phase, \emph{Data Analysis}, data are examined by data analysts.

Let us assume that  a flooding event has to be analyzed. 
As a first step, social media data are prepared in the \emph{Data preparation} phase.
%, which is then followed by a \emph{Data Analysis} phase.
To prepare the data, first we need to retrieve posts from social media, selecting the appropriate query for the search. The selection of keywords is critical, as on the one hand we aim to collect as much information as possible, whereas using generic keywords for events might result in a large amount of irrelevant content.
In some approaches, the native geolocations of posts are used for filtering the data. However, the proportion of natively geolocated tweets is very low;
%(estimates are around 0.3\% to 3\%), 
hence there is a high risk of excluding useful information if only the geolocated tweets are considered. 

The second step includes the removal of duplicates, basic cleaning actions on the raw dataset, for example, on poorly formatted text, and filtering posts to retain only those that are possibly relevant to the event.
One point of attention is that quality issues are often present when extracting information from social media. Texts and images might 
not be relevant to the event, and biases and fake news may be present.
In addition, the raw datasets resulting from the data collection step contain ``noisy'' data, including uninformative posts talking about the event or marginally related or generic texts or images, such as posts with memes. Therefore, filtering mechanisms must be put in place to prepare the dataset, clean the data (e.g., excluding exact duplicates and near duplicates), and  filtering only potentially relevant data. Several models for image and text classification in social media have been developed, mainly based on deep neural networks (e.g., \cite{asif2021automatic}). One of the open problems, beyond the development of the classifiers themselves, is the selection and configuration of the classifiers that are better suited for a given situation, as events can present distinctive characteristics in different contexts and locations around the world. As a result, the ``Cleaning and filtering phase'' is usually a rather complicated one, which is often developed specifically for a given data analysis project. 

The third step is an automatic annotation to enrich with additional  information derived from the automatic analysis of posts. Automatic annotation can include both labeling through automatic classification, for example,  the impact level of a naturale disaster in terms of visible damage in an image, and also automatic geolocation of posts that are not natively geolocated, adding location information.
This information is critical for identifying the posts  retrieved for the events to be studied.

At this point, the prepared dataset can be examined by the data analyst, within the \emph{Data Analysis} phase, in which hypotheses are explored (e.g, determining whether a given area was actually affected by an event or not) and information can be extracted. To support this analysis, crowdsourcing can be leveraged to provide additional contributions from citizen scientists, for example, by checking and refining the output of the automatic annotation and cleaning performed in the previous steps. As crowdsourcing activities can involve tasks that are inherently difficult (e.g., identifying if there are persons in a blurred image from a distance) or might be performed by an unspecialized crowd, a systematic analysis of the crowdsourcing-derived information must be performed to assess its validity. After the pre-analysis phase of the crowdsourced data, an actual data analysis can be performed to derive valuable knowledge. Different types of analysis can be envisioned, and the following are considered in this study:
\begin{itemize}
    \item \emph{Images of events}: the intended result consists in %proving the possibility of
    retrieving images from a location, usually displaying them using a GIS (Geographical Information System).
    \item \emph{Thematic mapping}: evidence, such as  the intensity of a flood event emerging from the prepared posts, is aggregated in given geographical areas, normalized and often weighted (e.g., based on population information). %characteristics, such as the size of the resident population%
    \item \emph{Alerts}: early alerts for a starting event are generated, and timeliness is the primary goal of the analysis. In this case, the balance between the manual and automatic activities to be performed must be assessed, as the results are useful if they are delivered in the first hours after the event onset \cite{anjum2021exploring}.
\end{itemize}
A more advanced type of analysis would be to solicit and then tackle problems proposed by the citizens themselves via crowdsourcing. This requires the ability to organize and analyze proposals, that can be provided by decision support tools, such as for instance Decidim4CS\footnote{\url{https://decidim4cs.ml}}. We do not discuss these tools further in this study, which mainly focuses on social media analysis.

Some critical issues emerge from projects studying emergencies (such as Crowd4SDG\footnote{\url{https://crowd4sdg.eu/}} and E2mC \cite{havas2017e2mc}), as well as from the literature (see Section \ref{sec:sota}):
\begin{itemize}
\item
Depending on the type of task being performed, \emph{time constraints} can be more or less stringent. In particular, there is a significant difference between alerts, which must be almost immediate to be useful, and thematic mappings, which are usually expected some time after the event.
In some cases, there can be little or no  time for human interventions in processing posts. 
Therefore, timeliness is  a parameter that should be considered when designing a system in this context.
\item
In addition, while the number of retrieved posts can be high,  relevant and useful posts may be few in number. Therefore, a sufficient level of recall is required during the filtering activities. Additionally, some circumstances or local situations may hinder the use of social media. Therefore, \emph{availability of data} is often  problematic. Moreover, the recall of relevant posts and the specificity of the collected data must usually be balanced for practical purposes.
\item
Finally, \emph{usability} of the tools is also an important aspect. Tools should be designed keeping the non-IT final users in mind. They should be very flexible in their configuration and provide user-friendly interfaces and interchange formats.
\end{itemize}
In the following sections, the Crowd4SDG approach for providing solutions to these problems is presented.
\vskip 30pt

\section{Citizen Science Solution Kit}
\label{sec:cssk}
\subsection{Crowd4SDG approach}
The varied requirements posed by different data analysis goals and circumstances require a flexible management of tools to support the collection, preparation, and analysis of data. In this section, we describe  \emph{Crowd4SDG Citizen Science Solution Kit}\footnote{\url{https://crowd4sdg.github.io/}} (CSSK) developed for this purpose, focusing on the components developed for social media analysis and crowdsourcing.

\begin{figure}
    \centering
    \includegraphics[width=0.9\linewidth]{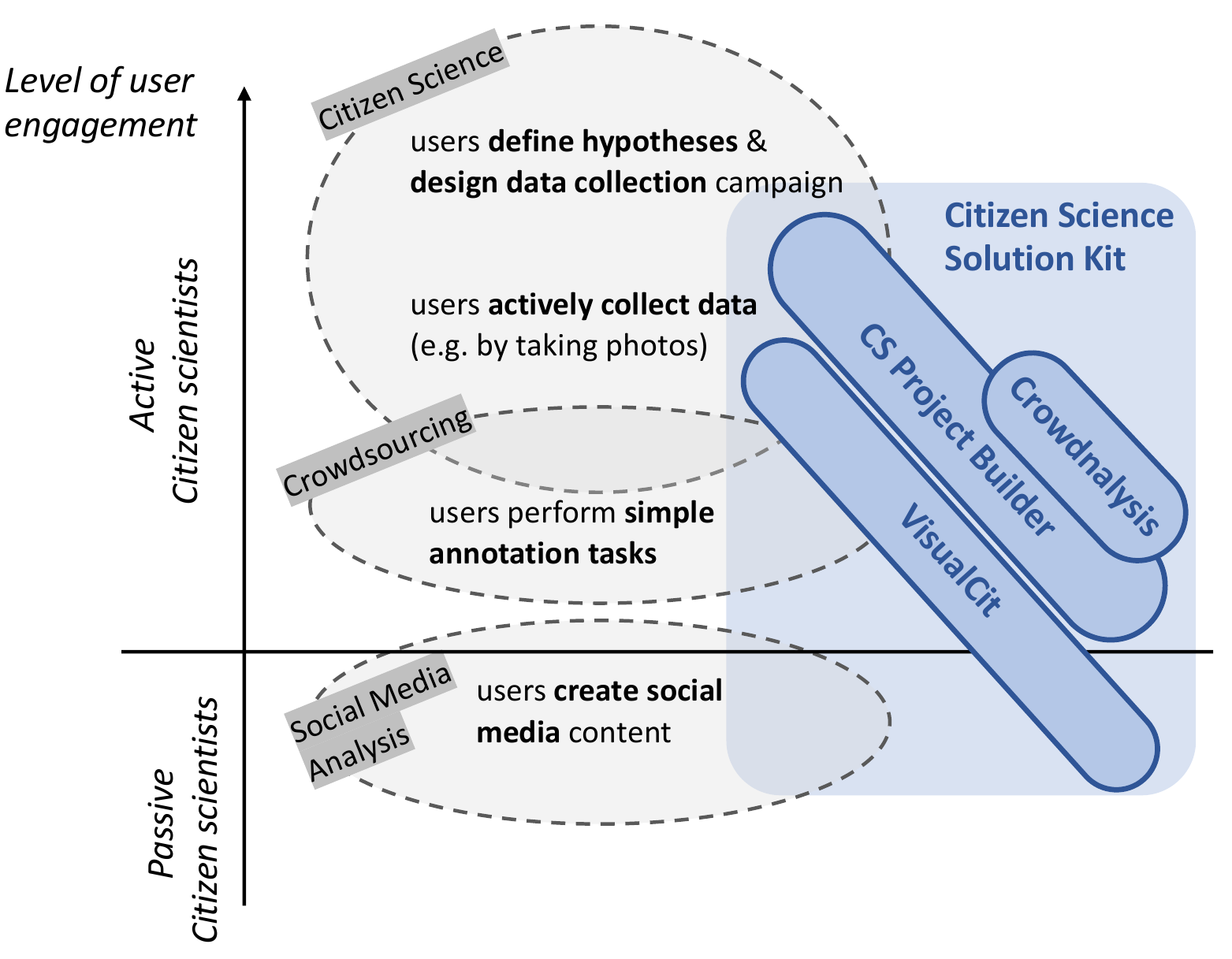}
    \caption{Citizen Science Solution Kit.}
    \label{fig:cssk}
\end{figure}

One of the key elements to be considered is the engagement of citizens in providing support, processing information, and suggesting solutions.
A possible analysis framework was proposed by \cite{data-democracy}, in which candidate tools and initiatives are classified according to both  the level of required engagement of citizens and  the implicit/explicit nature of the geographical information.
As shown in Fig.~\ref{fig:cssk}, we focus on building solutions for the different levels of citizens engagement.

%\begin{figure}[h]

With the \emph{VisualCit} (Visual Citizen) tool, we focus on collecting information from passive, citizen-generated content, starting however from the active participation of citizens in defining the type of information to be collected. VisualCit can also be used  as an active collection tool, in cases where specific hashtags focusing on a given event or situation are used, as proposed for instance in \cite{grasso2016codified}. The goal of VisualCit is to enable citizens and data analysts to query social media in order to obtain an input dataset, and to interactively define the filters and annotations to be applied to this dataset.

The \emph{Citizen Science Project Builder} (CSPB or CS Project Builder) tool allows configuring questions to be asked to citizens (also denoted as crowd workers or simply workers in the following) and gathering their answers.
These questions can be about the evidence collected from citizen-generated content on social media.
Questions can be of different types, ranging from confirmation of the relevance of selected images, to finding information in texts or images, or contributing new information on a given situation as free text or by assigning a label. %The final goal is to crowdsource information from citizens.

As crowdsourced information is known to be subject to possible biases or errors by the contributors, the same questions are asked to more than one citizen (usually between 3 and 10), and an analysis of the behaviors of the crowd must follow, to identify possible problems and to finally collect the citizens' contribution with a wisdom of the crowd approach. The analysis of crowdsourced information is performed with the \emph{Crowdnalysis} tool, thus providing aggregated Citizen Science results.

\begin{figure}
    \centering
    \includegraphics[width=1.0\linewidth]{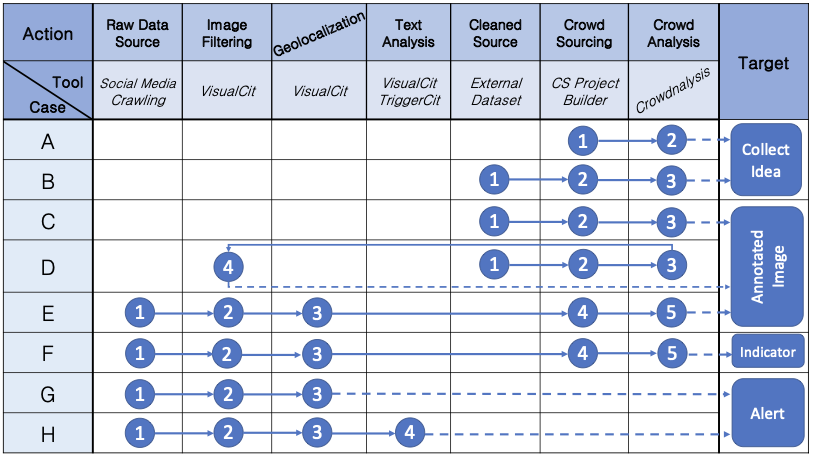}
    \caption{Use cases for the Citizen Science Solution Kit.}
    \label{fig:methods}
\end{figure}

The tools described above are flexible and they are intended to be combined in various  ways. Fig.~\ref{fig:methods} illustrates the typical use cases of the tools, corresponding to the requirements and scenarios discussed in Section \ref{sec:scenario}.
In the figure, we show possible combinations of the CSSK tools that have been experimented. 
The use cases range from a direct collection of ideas from citizens through the CS Project Builder, starting from scratch (A) or from a set of posts to be analyzed (B), to a full-fledged data preparation and analysis pipeline  to clean and annotate posts (E) or to derive indicators or thematics maps (F) with the support of VisualCit components for search, filtering, and annotating images with geolocations. In all cases in which crowdsourcing is used, the Crowdnalysis tool enhances the analysis of crowdsourced information.
Use cases in which the analysis does not include crowdsourcing (G, H) are intended for cases in which timeliness is critical (producing Alerts) and might  or might not include the VisualCit text analysis components.
A further use case is shown in (D), where crowdsourced information is filtered using VisualCit after crowdsourcing and crowd analysis.

In Section \ref{sec:case-studies}, we  discuss  in more 
detail the validation of the Crowd4SDG method, focusing on large case studies following use cases  typologies D, E, G and H, which have been developed and validated extensively in the project.

\subsection{Architecture} \label{architecture}

\begin{figure*}
    \centering
    \includegraphics[width=\textwidth]{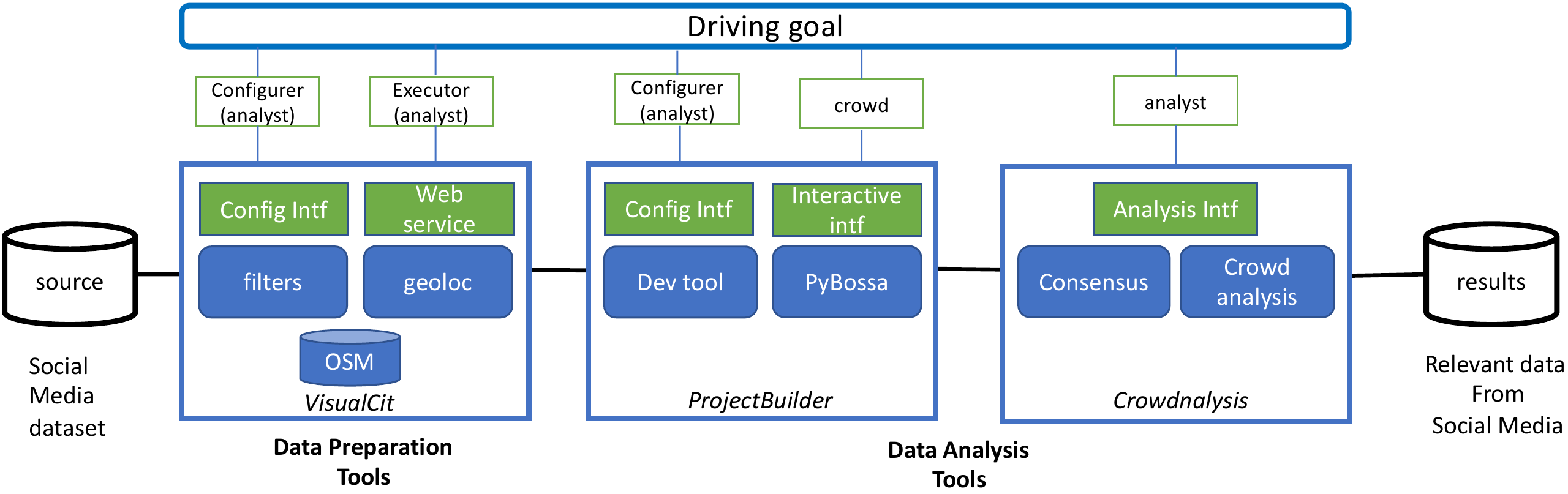}
    \caption{CSSK architecture.}
    \label{fig:architecture}
\end{figure*}

The high-level conceptual architecture of the Citizen Science Solution Kit is illustrated in Fig. \ref{fig:architecture}.

\begin{comment}
The CSSK provides an integrated framework for social media analysis, based on three main components: VisualCit, CS Project Builder, and Crowdnalysis.
%CSSK is supporting the preparation pipeline from raw data collection, transforming 
We regard CSSK as a pipeline that progressively transforms
a raw dataset $D_R$ into a prepared (i.e., cleaned, processed, annotated) dataset $D$ ready for data analysis. 
%More formally, the prepared dataset consists in a relation $D$ such as:

%\vspace*{\fill} 
%\begin{center}
%$\langle p,g,a\rangle \in D {\subseteq}\{P{\times}G{\times}A\}, P {\subseteq} R_D$

%$p \in P,g \in G,a \in A$}
%\end{center}
%\vspace*{\fill}

The prepared dataset consists on a set $P$ of relevant posts  cleaned from irrelevant information; a set $G$ with annotations $g$ about geographical information; and a set $A$ with task annotations performed during crowdsourcing. CSSK components are responsible for subsetting and populating these sets, as well as determining the relations between them.
\end{comment}

The CSSK provides an integrated framework for social media analysis based on three main components: VisualCit, CS Project Builder, and Crowdnalysis.
We regard CSSK as a pipeline that progressively transforms a raw dataset $R_D$ of collected social media information into a prepared---i.e., cleaned, processed and annotated---dataset $D$ ready for data analysis. The prepared dataset consists of: a set $P$ of relevant posts cleaned from irrelevant information; a set $G$ with annotations regarding geographical information; and a set $A$ with task annotations performed during crowdsourcing. CSSK components are responsible for subsetting and populating these sets, as well as determining the relations between them.
\begin{comment}
More formally, where $P{\subseteq}R_D$ is the set of relevant posts $p$ cleaned from irrelevant information, $G$ is the set of all possible geographical annotations $g$, and $A$ is the set of all possible task annotations $a$ that can be performed during crowdsourcing; the CSSK pipeline consists in a relation $K$ that relates $(p, g, a)$ to a data instance $d\in D$:

\vspace*{\fill} 
\begin{center}
$\langle (p,g,a), d\rangle \in K \subseteq \{\{P{\times}G{\times}A\}{\times}D\}$
\end{center}
\vspace*{\fill}

\end{comment}

As mentioned in the previous section, not all use-case scenarios involve  geolocation and crowdsourcing, therefore in this case the prepared dataset can also contain only a subset of the initial posts, only filtered and cleaned, with no geolocation nor additional annotations.
Each of the CSSK components is described in detail in the following sections. In the present section we provide an overview of its usage and of the main functionalities.

The CSSK components provide a framework intended to support \CS~(CS) projects, focusing on projects which are based on information retrieved from social media.
Citizen scientists can contribute to \CS~projects in many ways. For a detailed discussion, an interested reader can refer to \cite{fritz2019citizen}. As mentioned above, in the present study  we consider both active and passive citizens participation. Purely passive participation in our context consists in providing information about ongoing events that are being posted on social media, and in this case the goal in the CSSK is to process this information without further citizens involvement. As mentioned in the scenarios described above, we also consider a more active participation from citizens.
Citizen scientists can contribute analyzing, validating, and annotating information, or contributing new ideas and material.

To clarify the roles of citizen scientists in the use of the CSSK, we distinguish the following main roles interacting in the platform during data preparation:
\begin{description}
    \item[CS Project Manager] can be a single citizen scientist or a small team putting forward an idea and its corresponding data analysis, or the manager can be involved in a larger project involving many actors. In our case, the CS project manager is responsible for defining and refining the \emph{goal} of the project, the type of analysis to be performed and the type of results that must be provided.
    \item[Crowd] is composed of crowd {\bf workers}, that is,  citizens involved in providing information or annotations useful for the analysis goals. As described in Section~\ref{sec:Crowdnalysis}, different types of crowd workers are considered.
\end{description}

These three components have different functionalities. VisualCit is a framework that offers selection, cleaning and filtering components to be applied in order to reduce the number of irrelevant posts to be manually analyzed. The main goal is to improve the overall quality of the dataset. VisualCit also provides  automatic geolocation functionalities, based on OpenStreeMap \cite{DBLP:journals/pervasive/HaklayW08} and the CIME approach \cite{scalia2021cime} for geocoding. The CS Project Builder provides an environment for crowdsourcing based on the open-source  PyBossa framework. Crowdnalysis analyzes the quality of  crowdsourcing and crowd behavior.

CS project managers can play three different roles:
\begin{itemize}
\item \emph{Configurer}: When CS Project Managers have set their goals, they must configure the data preparation pipeline, selecting  the analysis scenario, the volume of data to be analyzed and the frequency of the analysis. In this phase, the tools are configured, along with their parameters. If crowdsourcing is part of the scenario, the crowd and the crowdsourcing tasks for annotations are identified by CS Project Managers.
\item \emph{Executor}: Once the data preparation pipeline has been configured, the CS Project Manager can execute it on raw data as needed. In this phase the intermediate results are also assessed, which might result in a reconfiguration of some of the tools if the quality is not satisfactory, or even a revision of the initial goal. % of the project.
\item \emph{Analyst}: After crowdsourcing, the CS Project Managers analyze the results of the crowd. In this phase, they need to identify the most appropriate crowdsourcing analysis model   for computing Consensus and they should also analyze the behavior of the crowd (crowd analysis), looking for anomalous or opportunistic behavior that might invalidate the results. Considering the obtained results, quality-improving actions can be performed.
\end{itemize}

%As mentioned above, we distinguish from configuration and analysis phases, and execution phases.
In general, a
visual interface is provided to support configuration operations on the VisualCit and CSPB tools.
In VisualCit, execution can be performed using an interactive interface for the initial samples, whereas execution on large volumes of data is performed using a service-based architecture.
In CSPB, an interactive interface is provided for execution of the crowdsourcing tasks by the crowd and to extract data for analysis by the Project Manager.
Therefore, both components provide a set of web services that can be invoked via APIs from both interactive and programmatic interfaces.
Once the data are prepared using VisualCit, crowdsourcing can be performed. A common data exchange format is defined for the three tools based on \csv~files.
%Each tool can be invoked by the user with the interactive interface and through web services. An interactive interface is provided to the crowd to perform the tasks.

%\vskip 1cm

%think: decidim4cs. a usage story without a case study?.. 

\subsection{Selecting and filtering data: VisualCit} 
\label{sec:visualcit}

\subsubsection{Filtering}

VisualCit is a social media processing tool aimed at exploring and building data processing pipelines. It is meant to be general purpose, pluggable to different data sources, and easily extensible.
% \Figure[t!](topskip=0pt, botskip=0pt, midskip=0pt)[width=1.5\columnwidth]{figures/visualcit.pdf}
% {VisualCit interactive interface.\label{fig:visualcit}}

\begin{figure*}[th]
    \centering
    \includegraphics[width=\textwidth]{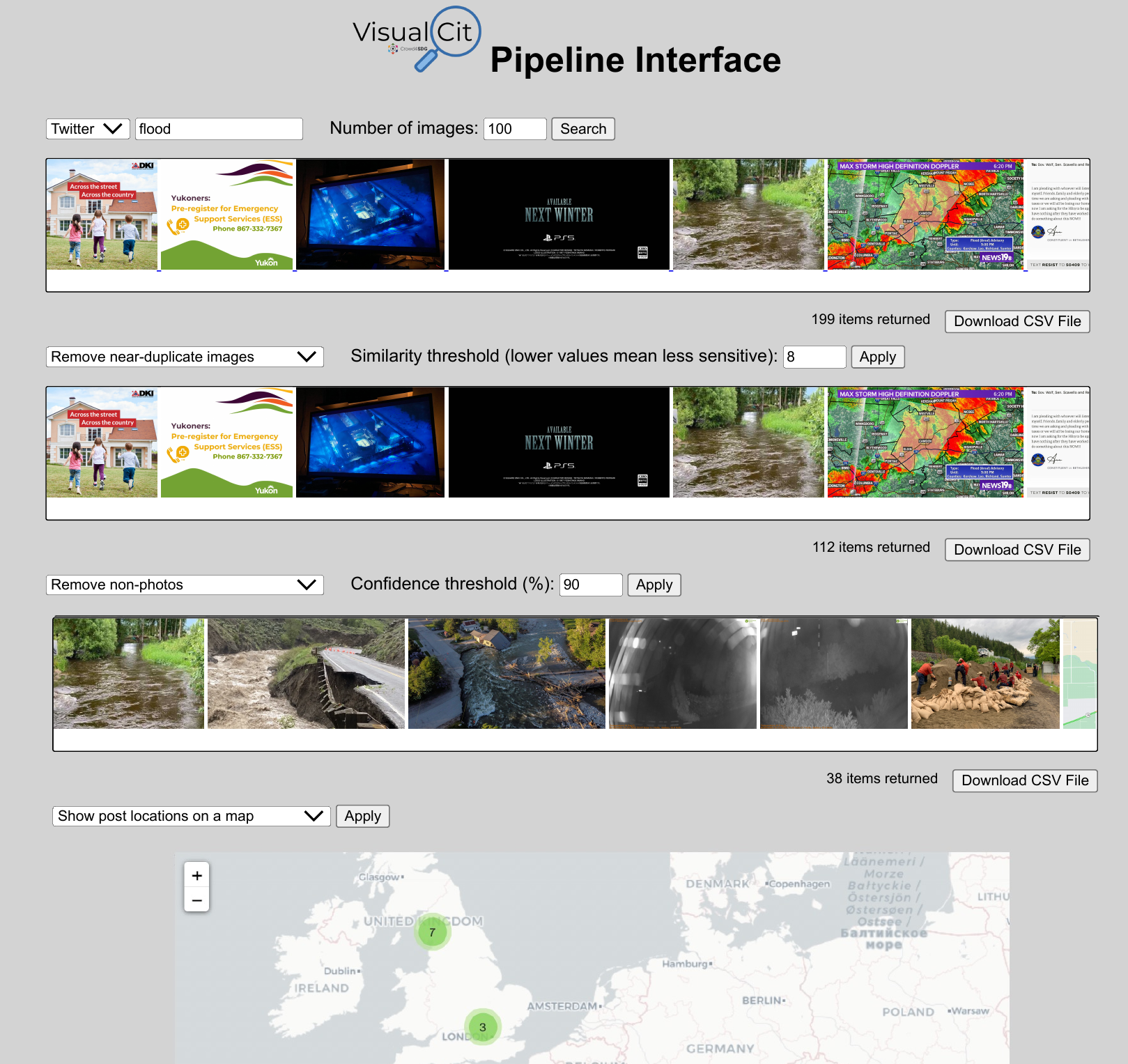}
    \caption{VisualCit interactive interface.}
    \label{fig:visualcit}
\end{figure*}

\begin{figure}[h]
    \centering
    \includegraphics[width=0.9\linewidth]{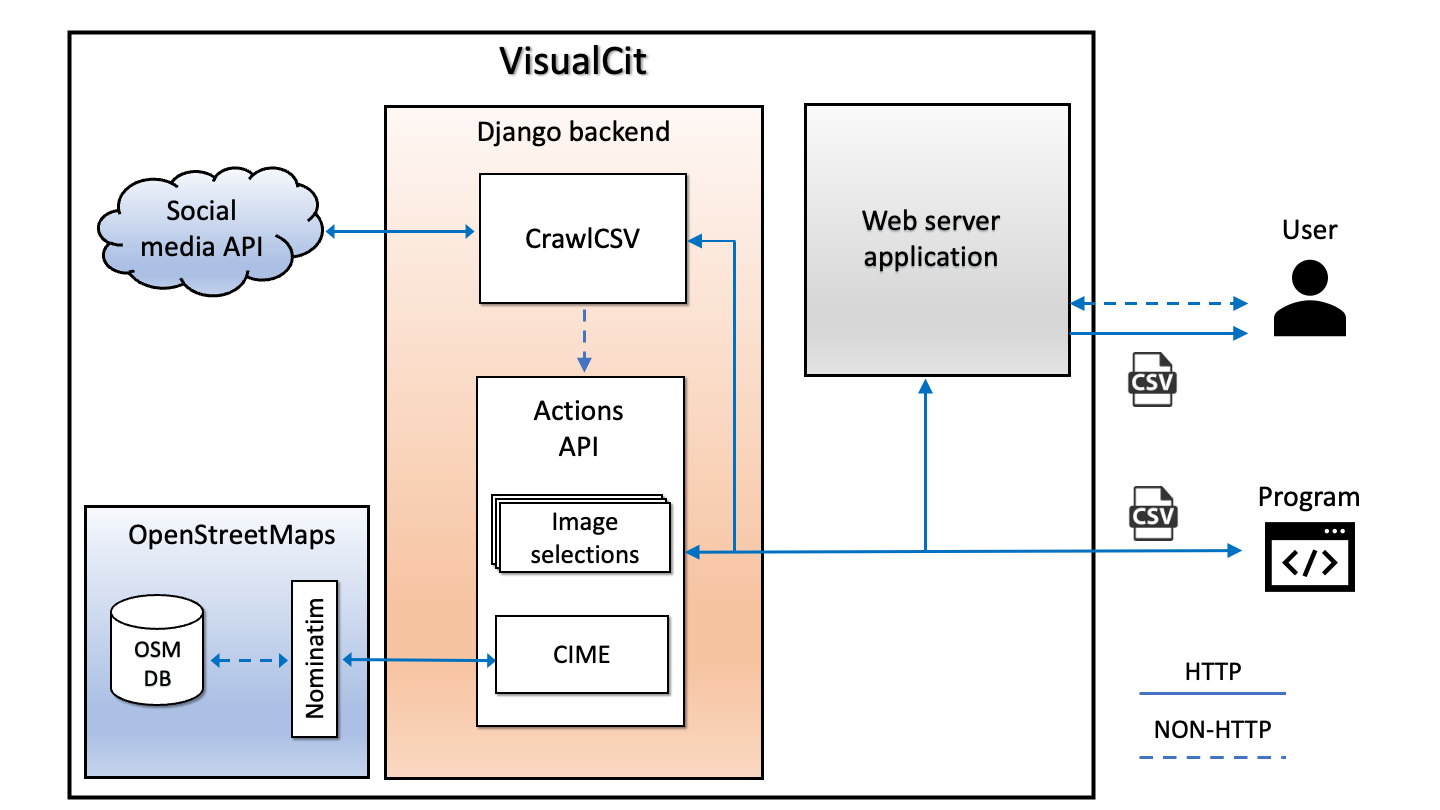}
    \caption{VisualCit architecture for selecting filters and geolocations annotation.}
    \label{fig:visualCit-architecture}
\end{figure}

VisualCit functionalities are available in two ways. The first is an interactive web interface (see Fig. \ref{fig:visualcit}), which is useful for exploring data and checking the results of applying actions to the data. This tool can be used to quickly sketch data processing pipelines and adjust parameters with immediate visual feedback.\footnote{A demo of the web interface is available at \url{http://visualcit.polimi.it:7778}} The second mode is an HTTP web service that exposes the GET and POST methods. The web service accepts single processing requests or composite pipeline configurations, executes them, and sends the results back. A high-level overview of the architecture is presented in Fig. \ref{fig:visualCit-architecture}.

At its core, VisualCit exposes a set of configurable data processing actions. The action library is extensible and is mostly aimed at actions that can be applied to media. Supported media are currently images. The classes of data processing actions currently supported by VisualCit are
i) ingestion, which collects data from social media with configurable criteria; ii)
cleaning, mainly to reduce the number of duplicate posts and images; iii)
selection, to select posts that are relevant to the goals of the social media data analysis to be performed; and iv)
augmentation, to automatically derive further information on the posts to be analyzed.

Actions may have one or more configurable parameters, such as confidence thresholds. These parameters can be provided from any of the two interfaces. If they are not provided, configuration defaults are loaded. In the web interface, parameters can be configured via form fields, whereas in the web service they can be provided as key/value pairs, where keys correspond to the parameter names and values to the parameter values. Key/value pairs can be provided as HTTP request parameters, or alternatively as a JSON payload. The web service also accepts multiple actions to be executed together as a pipeline. This is a natural extension that enables the execution of an ordered list of actions, which can be regarded as a composite transformation or a data preparation pipeline. Currently, in the front-end interface, there is a prototype for exporting a processing configuration that is built interactively. This configuration can be sorted, edited and sent directly to the web service backend, for large-scale and batch processing.

The data interchange format, for both input and output, is based on the \csv~format. The minimal set of attributes to be provided consists only of a media URL attribute, referencing the media to be analyzed. 
%Usually, a unique identifier of a social media post is provided, together with a string representation of the textual content of the post itself. Additional metadata can be present in the file and propagated transparently by VisualCit. Each action has a set of output attributes that are added to the data when processing occurs. Records are removed from the dataset when certain actions determine that the corresponding item should be filtered out.

Data can be provided to VisualCit explicitly as records in a \csv~file. The data to be processed can also be ingested directly from social media. An adapter for extracting social media posts should be implemented for each compatible platform. Crawling is then configured through a query string, accounting for keywords, key phrases and possibly advanced search operators. Options such as date and location filters are also possible depending on their availability on social media platforms. The adapter accounts for paginating the results and building a unique dataset, which can then be processed. Crawling can be invoked as any other action and executed before any action. In this way, querying the data source and processing the results with a pipeline can be performed by providing a single configuration. If no crawling action is provided as a starting point, the data to be elaborated must be sent as a POST payload.

For efficiency reasons, the media content is cached on local storage upon the first request. This is usually performed when the first action is executed on an item. Moreover, data can be processed using a configurable level of parallelism. This is done to take advantage of the possible underlying hardware architectures. For CPU computing, parallelism was obtained through an asynchronous thread pool of workers. For GPU computing, if enabled in configuration, the parallelism level represents the number of items concurrently loaded in the GPU memory. This is especially convenient when actions involve the use of certain types of deep neural networks, such as convolutional neural networks (CNNs), which can take advantage of such a setup.

Among cleaning actions, VisualCit implements deduplication functionalities. Deduplication can be performed against 1) identical URLs, to avoid analyzing duplicated links, 2) exactly matching images, through standard hashing techniques, and 3) almost identical images, using a perceptual hashing algorithm, with a configurable similarity threshold.

Selection actions, or filtering actions, generally leverage CNNs to enable queries on the contents of image data to perform selection at a semantic level. The application of this kind of approach has been the object of study in the field, for example, in \cite{nguyen2017automatic} and \cite{DBLP:conf/icse/NegriSARSSFCP21}. Some of the available classifiers are off-the-shelf whereas others are custom. In VisualCit, state-of-the-art object classifiers are available\footnote{Currently, DETR and YOLOv5} so that queries can be made about the presence of an object or a number of objects in the image. Depending on the operational requirements, a confidence level can be provided to accept or reject the presence of an object with customized sensitivity. A scene classifier is also available\footnote{PlacesCNN, trained on the Places365 database}, enabling queries about specific kinds of scenes or aggregated scenes such as indoor/outdoor spaces or private/public spaces. An action for removing NSFW content is provided.\footnote{NSFW means ``not suitable for viewing at most places of employment'' according to the Merriam-Webster online dictionary} An action for detecting and filtering non-photographic content is also provided in order to remove images corresponding to drawings, screenshots, memes and other types of non-photographic content. This is usually helpful when looking for visual evidence during emergency events. Finally, a custom flood classifier can be used as an example of an event-specific classifier. This classifier attempts to estimate whether the images depict a flood event.

VisualCit is not limited to data filtering or selection. Augmentation actions are also possible, for example, by enriching post data with external data sources. A notable example is geolocation, through which geographical positions are automatically associated with social media posts. This is usually critical for obtaining a spatial representation of an event. Geolocalization in VisualCit was performed using the CIME algorithm, which is detailed in the next subsection.

\subsubsection{Geotagging}\label{subsec:cime}
To provide evidence that can be used in an emergency response, the selected content is not useful unless associated with a position. For example, a picture could depict a specific place or be associated with a wider area (e.g., a city), and there is a need to associate this information with the original data with some confidence. As noted, the number of social media posts that are natively associated with a location is usually small. Posts may also contain other geographical references, such as in textual form, that can be exploited to associate possible locations with a post. Candidate locations are extracted from the text using a multilingual named entity recognition (NER) library\footnote{Polyglot \url{https://polyglot.readthedocs.io/en/latest/}}. The NER functionality extracts strings that could refer to some kind of entity, in our case, a geographical location. The language used for the analysis of the posts could be provided externally using a language detection algorithm, propagating known information about the language (e.g., coming from the social media platform), or automatically inferred by CIME. Disambiguation of the candidates must then be performed to isolate meaningful candidates. This is usually a challenging task because the same entity could be referred to with many names, and different places could share a name, or parts of it. To link candidate strings to location entities, Nominatim\footnote{\url{https://nominatim.org/}} was used as a gazetteer. Nominatim leverages OpenStreetMap data to link geographical entities to the detected candidates. After disambiguation, the set of most promising results was returned. Further details on the operation of CIME can be found in \cite{scalia2021cime}. Further priors on the location of the post could also be leveraged to enhance the results. For example, information about the language of the post can be used together with associated administrative boundaries to exclude some of the results.

\subsubsection{Keyword enrichment mechanisms}
\begin{figure}[h!]
\centering
  \includegraphics[width=0.9\columnwidth]{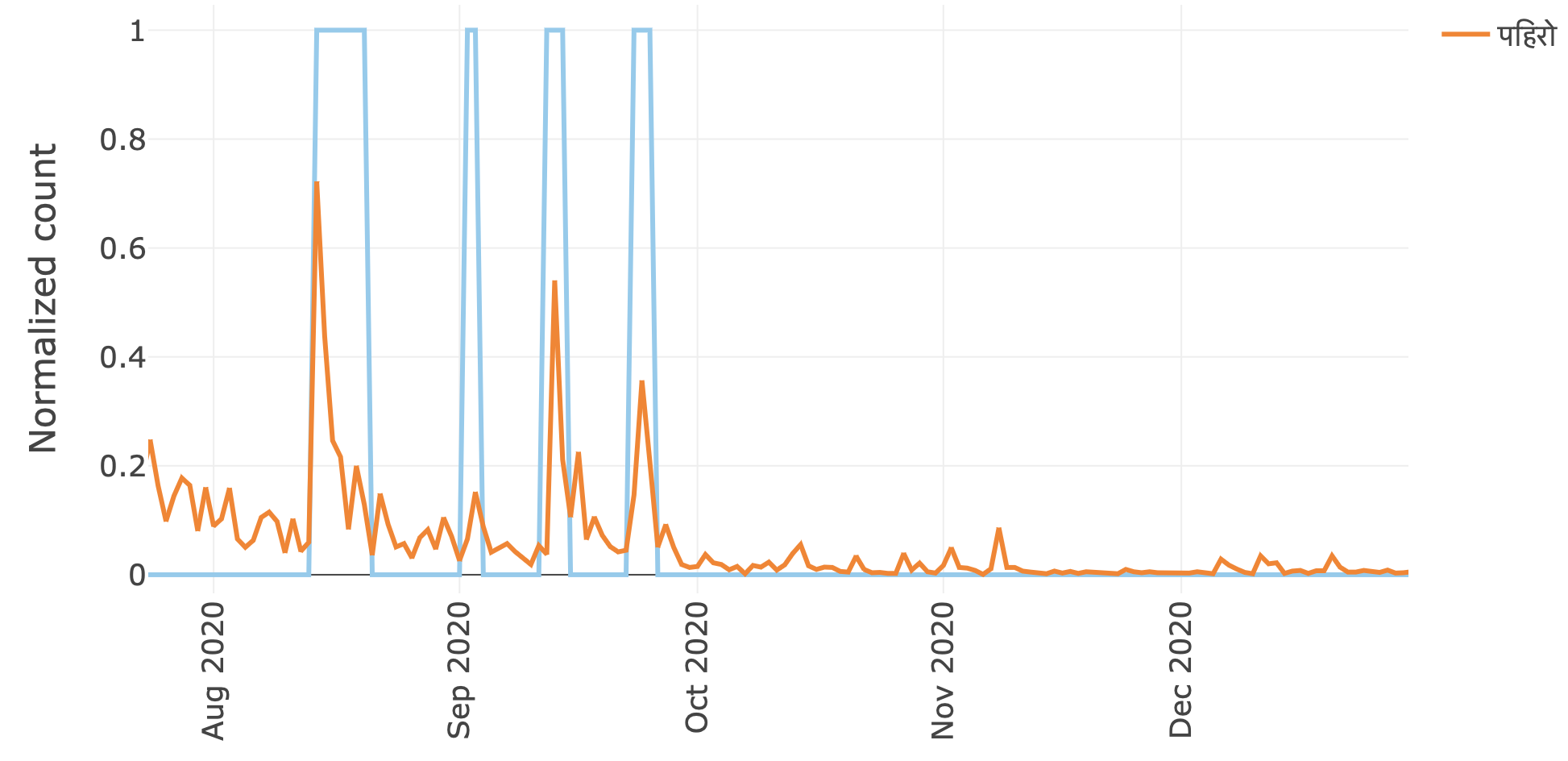}
  \caption{Occurrence of the term "landslide" compared with flood events reported by GDACS in Nepal.}
   \label{fig:landslide}
\end{figure}

Using VisualCit, we focus on visual evidence that can support an emergency response. This is somewhat complementary to text-based approaches that have wide applications in the field. In some tasks, the textual dimension is indispensable. This is the case, for example, in building queries that are used to ingest data from social media platforms. In \cite{bono2022triggercit}, we focused on light dictionaries to recognize the onset of natural disaster events. An appropriate dictionary can enhance both the recall and precision of the collected data, as discussed in Section \ref{sec:scenario}. We then studied how to automatically extract a relevant dictionary when a ground truth about the events of interest was available. An intuition of the approach is illustrated in Fig.~\ref{fig:landslide}.

By sampling posts from days in which the events occur, together with days in which they are not, a measure of the correlation between keywords and events can be established. We used time-lagged cross-correlation between the word counts and event onsets to measure the intensity of their relation. A one-sided variant was used to avoid inverse causality. The correlation measures are usually weak but significant in terms of recall augmentation. A manual filtering step, coupled with automatic translation, was also involved. Filtering was implemented using checkboxes. This method enables an effortless construction of language-based dictionaries.

At the intersection of image-based and text-based approaches, the available post data with filtered images can be used to augment the results by looking at posts with no attached images. A straightforward approach is to extract tweets that are similar to those selected by VisualCit, which are generally associated with a better relevance to the goal. Textual frequency analysis can then be used to identify keywords that were not initially considered. These new keywords can be used to select other posts, thus obtaining a  wider result set (e.g., posts that were not fed to VisualCit because they had no media attached), and extract a new dataset from social media, whose unseen items can then be fed to VisualCit. This approach can be used to increase the number of locations extracted from a dataset. Textual similarity scores were used to rank the correspondence of the posts to those filtered by VisualCit based on media content. Finally, the most promising posts were fed to CIME for geolocation. This can enhance the geographical description of the event as more data points are available. More details on how this approach can be used are provided in \cite{bono2022triggercit} where TriggerCit, a VisualCit extension aimed at generating alerts, was presented.

\subsection{Building Crowdsourcing projects with Citizen Science Project Builder} \label{sec:projectbuilder}

Depending on the domain and scientific discipline of application, \CS~has multiple practices associated with it. Recent literature \cite{Strasser2018} proposes a general classification based on the level of engagement and type of \CS~activity, namely:
\begin{itemize}
\item \emph{Volunteer sensing} - where participants use available sensors (e.g., in smartphones) to collect data that are then used by scientists for analysis. 
\item \emph{Volunteer computing} - a method in which participants share their unused computing resources on their personal computers, tablets, or smartphones and allows scientists to run complex computer models.
%when the device is not
\item \emph{Volunteer thinking} (or analyzing) - where participants contribute their ability to recognize patterns or analyze information that will then be used in a scientific project. 
\item \emph{Self-reporting} - where participants share and compare medical information as both qualitative (self-reported symptoms and illness-narratives) and quantitative data (patient records, genomic and other laboratory test results, and self-tracking health data). 
\item \emph{Making} - these practices are based on “making” things, often low-cost DIY sensors, and use them to collectively produce knowledge.
\end{itemize}

\begin{figure*}[h]
    \centering
    \includegraphics[width=\textwidth]{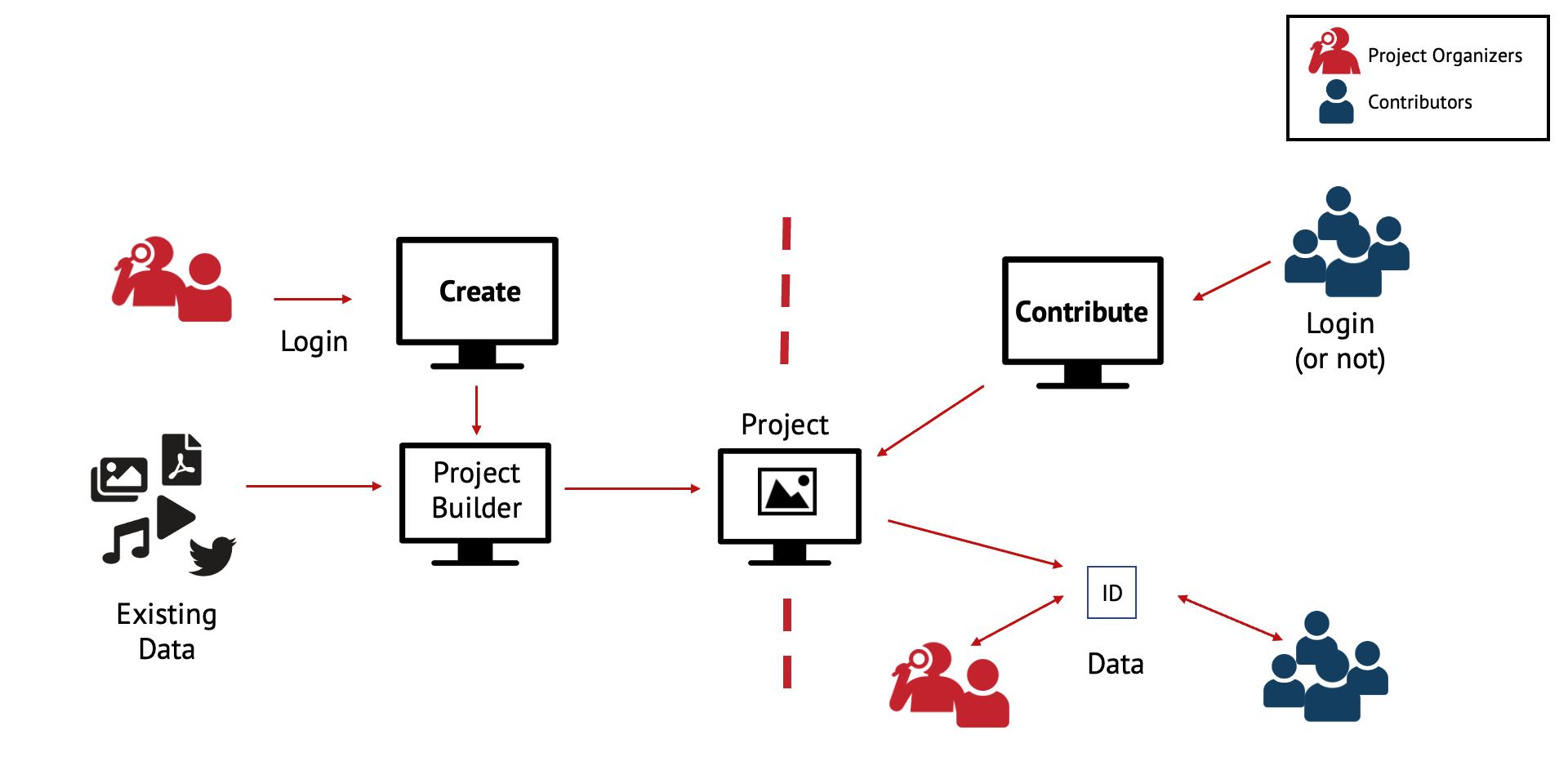}
    \caption{CS Project Builder overview.}
    \label{fig:CSPB_2}
\end{figure*}

The first three categories can also be grouped under the heading “citizen cyberscience”, a term created to describe activities that rely on the use of computers and the Internet \cite{Grey2009}.
The \emph{Citizen Science Project Builder} (CSPB) is a web-based tool that allows researchers, students, and all members of the public to create and run, and also participate in, volunteer thinking projects. Specifically, projects where volunteer contributors are asked to perform complex data classification tasks that are still best performed by human minds and skills, such as classify, tag, describe, or geo-localize existing digital data. Examples include classifying images of snakes, transcribing handwritten German dialect, or describing the content of video clips. In particular, the CSPB supports projects based on digital data in the form of images, text, \pdf~documents, social media posts (e.g., tweets), sounds and video clips.

\begin{figure*}[h]
    \centering
    \includegraphics[width=\textwidth]{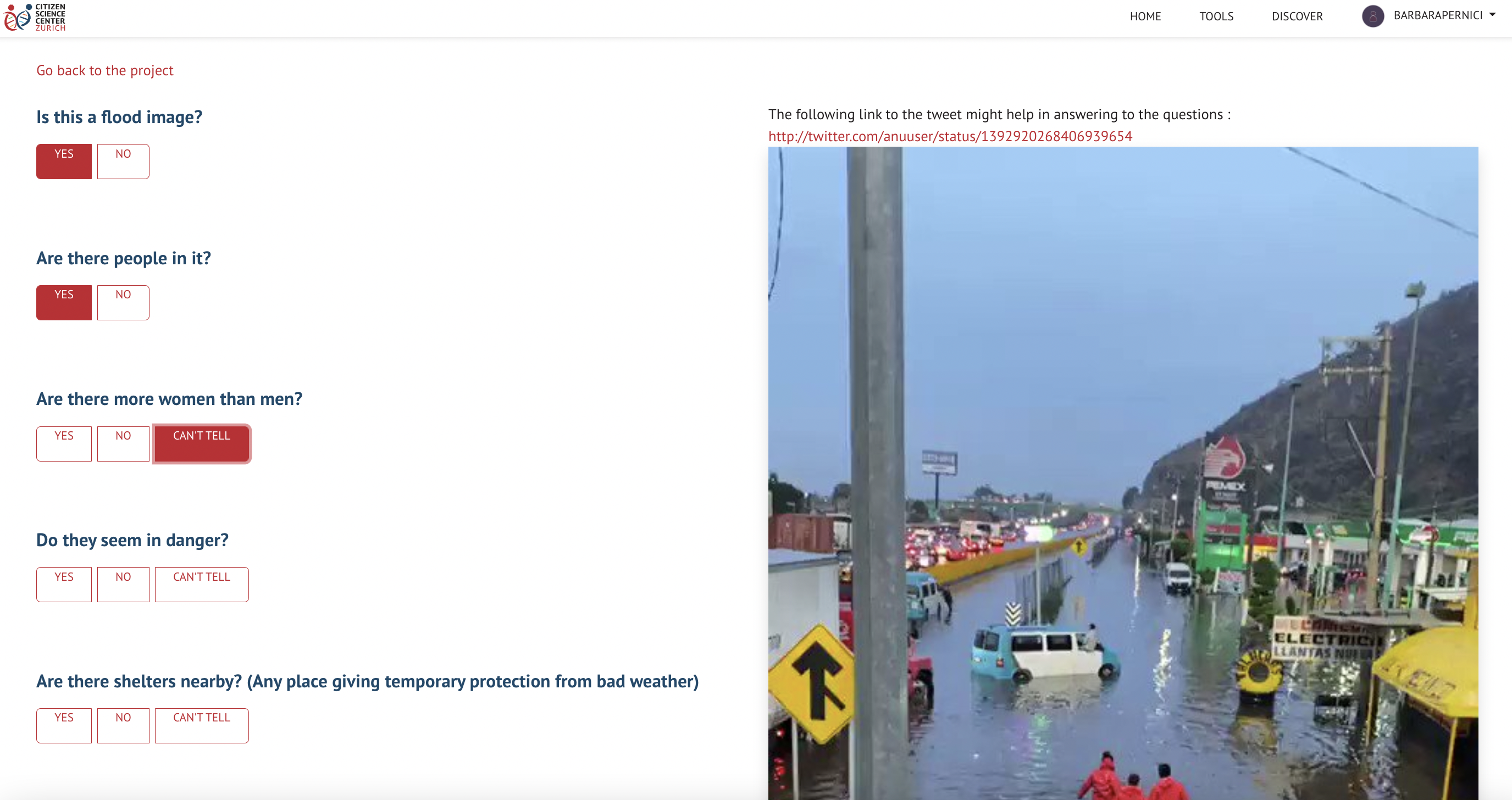}
    \caption{Example of CS Project Builder crowdsourcing interface.}
    \label{fig:cspb}
\end{figure*}

In practice, the CSPB is a web interface that provides access to a dashboard where projects can be set up following a simple step-by-step process. Each step is clearly described in a text/image-based tutorial available on the platform, and the process can be summarized as follows (see Fig.~\ref{fig:CSPB_2}):
\begin{itemize}
\item \emph{Step 1: Project description} - Provide a nice and catchy title for the project, an image to represent it, and some additional text to explain in simple terms the purpose of the project, why it is important, how volunteers can contribute, how their contributions will be used, who is behind the project, and links to additional information available online.
\item \emph{Step 2: Data source and type of task} - Select the type of source files to work with. The options include: images (.jpg and .png), videos (.mp4), audios (.mp3), tweets, \pdf, and maps. Then, the task the contributors will need to do, including “survey” format for tasks such as classification, description, and counting, and “survey plus geolocation” if the task also includes geolocation on a map (e.g., identifying a landmark in an image and locating it in a city).
\item \emph{Step 3: Design your task protocol} - Design the survey protocol, including many options of questions and answers, or text fields, just as easy as a typical survey-building interface.
\item \emph{Step 4: Select the location of source files} - Import the source files (images, videos, sounds, etc.). The possible data sources include: Dropbox, Flickr, Amazon S3 buckets, and \csv~files. There are particular cases for Twitter, including files built with VisualCit, or for data collected with the CS Logger (another tool being developed as part of the CSSK). 
\item \emph{Step 5: Test and Publish} - From the project dashboard page, the draft can be modified and shared with collaborators for feedback and iterations. Several parameters of the project can be monitored and set from this dashboard, including statistics and tasks and data management. Once ready to go public, the project can be submitted for publication. The Administrators team will ensure that the project adheres to legal/ethical criteria before making it available for public contributions. Once the project has been published, anyone can participate in it. 
\end{itemize}

The entire process requires limited technical knowledge and little or no coding skills. However, the project dashboard also provides access to a “coding” interface where the project protocol code is accessible and modifiable. Users with Vue.js coding skills can act directly on the code to make any desired modifications to the interface.

Fig. \ref{fig:cspb} shows an example of the typical interface generated by the CSPB: the right side of the screen displays the digital data, in this case a tweet including both text and an image. The left part features the survey protocol, in this case a series of conditional questions about the content of the tweet and related images.

The CSPB implementation is based on the open-source crowdsourcing framework PyBossa and its code is publicly available under the ‘CitizenScienceCenter’ organization on GitHub. PyBossa\footnote{\url{https://pybossa.com/}} is an ongoing open-source development project. It is an extremely flexible and versatile technology used for the development of platforms and for data collection within collaborative environments, analysis and data enrichment. PyBossa is implemented with a RESTful API\footnote{\url{https://docs.pybossa.com/api/intro/}} to easily distribute tasks and collect, maintain, and process data from volunteer participants. In addition, the PyBossa RESTful API allows creating/removing projects, as well as the implementation of graphical interfaces. 

The PyBossa software architecture is composed of three major components (see also Fig. \ref{fig:crowdnalysis_service}):

\begin{itemize}
    \item The \emph{Task Importer} allows the creation of a new PyBossa project and tasks. This can be called from both the web interface and command line using the PyBossa API.
    \item The \emph{Task Presenter} eases the creation of the front end of the PyBossa projects. The task presenter accepts Vue.js code, which provides total flexibility to develop complex data analysis interfaces for audio, video, photos, or \pdf~documents. 
    \item The \emph{PyBossa Core} builds on a PostgreSQL database to store all information related to users and projects. The PyBossa Core handles the scheduler to allocate tasks to users, redundancy, and export options. It also handles the execution of webhooks to manage the execution of cron processes. 
\end{itemize}

The CSPB as well as PyBossa allows three different user roles:

\begin{itemize}
    \item The \emph{Project Organizer} is assigned as soon as the user creates a new project, and is related and limited to the projects that the user owns. The project Organizer has the right to edit the content of the project, update task visualization, add/delete tasks, export data and results. 
    \item The \emph{Contributor} is any user who participates in a CSPB project by analyzing data and submitting their results. Contributions can be assigned (after login) or anonymous.
    \item The \emph{Admin} role is reserved for the administration of the platform, and it gives similar privileges to project organizers over all projects. 
\end{itemize}

The CSPB builds on the PyBossa RESTful API, providing a graphical interface that exploits the potential of PyBossa without the need for interacting using bash or any programming language. 

\begin{comment}
\textbf{@Jose - can you see if in github you can find a schematic of the current backend?}
 Fig. \ref{fig:CSPB-architecture}
\end{comment}

%\begin{figure}[h]
%    \centering
%    \includegraphics[width=0.9\linewidth]{figures/CSPB_architecture.png}
%    \caption{PyBossa architecture}
%    \label{fig:CSPB-architecture}
%\end{figure}

\subsection{Social media consensus analysis with Crowdnalysis}
\label{sec:Crowdnalysis}

%State the need; link the need with CSPB; give a use case; reuse the Fig 1 in MDPI paper; TODO  briefly describe the three spaces and probabilistic model, maybe with Fig 2 in MDPI paper; reuse the text form Section 5.2 of D2.1---giving examples to which questions our probabilistic model can answer; remark it is implemented in Python and give the class diagram figure; describe implemented consensus models---including access to Stan implementations via CmdStan interface; remarks on the generative base class and prospective analysis; extend the abilities with question inter-dependencies, distinguishing real classes from reported labels, visualization; cite the PyPI package, test coverage, and the github repo with CI+CD workflows; mention automatic consensus computation in CSPB by Crowdnalysis-service (beta), give a use case with a figure as in D2.2; give future work---i) incorporating hierarchical models, ii) implementation of the remaining answers to above-mentioned questions, i.e., facilitating task$\leftrightarrow$annotator assignments; \cite{cerquides2021}including dependencies. (OM)

When the volunteers contributing to a \CS~project in CSPB complete the classification tasks (e.g., labelling the severity of damage in photos) on the platform, the individual annotations of the crowd members need to be aggregated to achieve a consensus on each task. Majority Vote is the most common aggregation method where the consensus on a task is the most voted class, and each annotator's vote has an equal contribution. This method counts on the wisdom of the crowd, but it cannot tackle situations such as  when i) we are unsure of how many annotators are required to obtain reliable results and ii) some annotators perform better on some tasks than others.
%; iii) there might be spam annotators randomly labelling the tasks, etc.
We developed the Crowdnalysis \cite{crowdnalysis2022} software library to address the critical needs that arise in the planning of a crowdsourcing project and the analysis of crowd-sourced data in CSPB.
Crowdnalysis incorporates probabilistic consensus models that estimate individual annotator error rates for a given set of tasks even when the ground truth is unavailable. These models enable Crowdnalysis to weigh the contribution of each annotator and thus, yield a more reliable consensus. 

\begin{figure}[!th]
    \centering
    \includegraphics[width=0.98\linewidth]{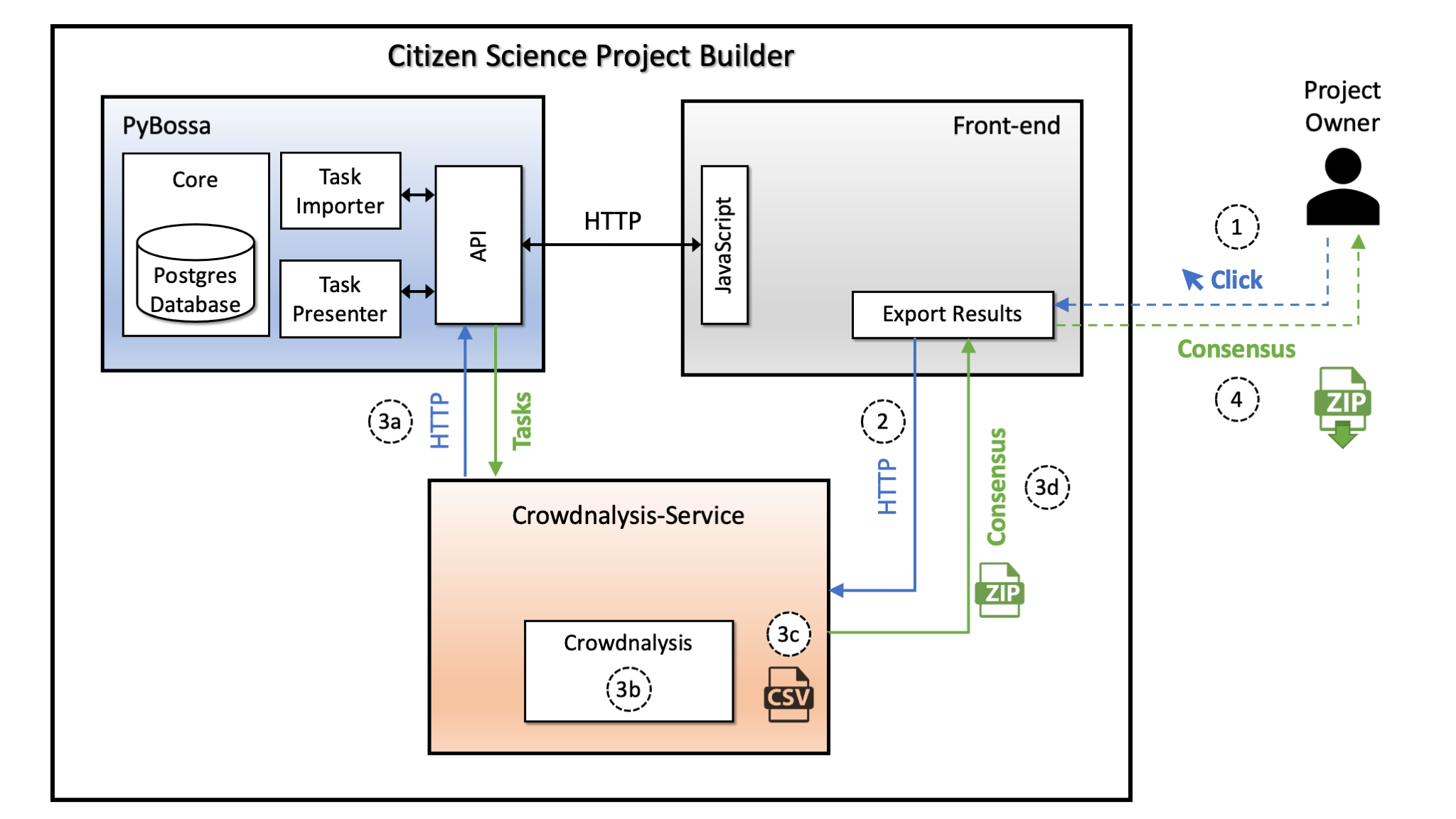}
    \caption{Automatic computation of consensus in CSPB via Crowdnalysis.}
    \label{fig:crowdnalysis_service}
\end{figure}

Crowdnalysis is integrated into CSPB as an on-demand service. Fig.~\ref{fig:crowdnalysis_service} illustrates the use case: 1) The project owner clicks the ``Export Results'' (of annotations) button on the front-end; 2) The request is forwarded to the Crowdnalysis-Service; 3) The service:
a) Calls PyBossa API to extract task and annotation data; b) Computes the consensus on tasks for each question asked to the crowd by using Crowdnalysis with the given consensus model; c) Creates a Comma-Separated Values (\csv) file for each consensus depending on the user's request; d) Sends the consensus and original result files back to the front-end in a \zip~archive file; 4) The user downloads the \zip~file without leaving the front-end in any of the above steps.

We introduced the conceptual and mathematical framework of Crowdnalysis in \cite{cerquides2021}. The main concepts (inspired by \cite{Dumitrache2018a}) in our mathematical model are as follows:

\begin{description}%\begin{IEEEdescription}[\IEEEsetlabelwidth{$\alpha\omega\pi\theta\mu$}\IEEEusemathlabelsep]
\item[Worker] is any of the participants in the annotation process.
\item[Task] is the minimal piece of work that can be assigned to a worker.
\item[Annotation] is the result of the processing of a task by a worker.
\end{description}

\begin{figure}[ht]
    \centering
    \includegraphics[width=0.6\linewidth]{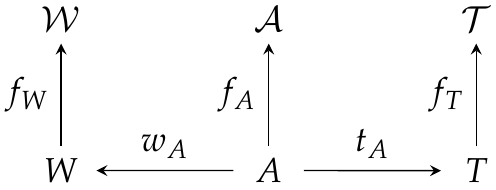}
    \caption{Spaces and mappings in Crowdnalysis' mathematical model.}
    \label{fig:crowdnalysis_map}
\end{figure}

Spaces and inter-mappings regarding these basic concepts are shown in Fig.~\ref{fig:crowdnalysis_map}. Workers, tasks and annotations are given by the finite sets of $W$, $T$ and $A$, respectively. Assuming that each worker, task and annotation can have distinctive features, we define the \emph{feature spaces} $\mathcal{W}$, $\mathcal{T}$ and $\mathcal{A}$, and \emph{feature mappings} $f_W\,:W\rightarrow\mathcal{W}$, $f_A\,:A\rightarrow\mathcal{A}$ and $f_T\,:T\rightarrow\mathcal{T}$ that altogether describe the individual characteristics of each member of the corresponding concepts. Finally, the functions $w_A:A\rightarrow W$ and $t_A:A\rightarrow T$ map each annotation to its worker and task, respectively.

From an epistemological point of view, we refer to crowdsourcing when we need to determine a specific characteristic(s)---which is(are) unknown to us---of each item in a set of tasks with the aid of workers annotating them for us. Accordingly, we assume that we can factor the task feature space as $\mathcal{T} = \mathcal{T}_O\times\mathcal{T}_C\times\mathcal{T}_H$, where $\mathcal{T}_O$ contains the observable task characteristics, $\mathcal{T}_C$ contains the unobservable (i.e., latent) characteristics in which we are interested in for the consensus and $\mathcal{T}_H$ contains the characteristics that are unobservable and we are not interested in. We model the lack of knowledge by means of a probability distribution, and subsequently aim to determine a probability distribution over $\mathcal{T}_C\times\overset{\mathtt{t}}{\ldots}\times\mathcal{T}_C$ where $\mathtt{t}$ denotes the number of tasks. We call this distribution a \emph{joint consensus}.

With the above conceptual framework, in \cite{cerquides2021}, we defined the consensus problem, introduced the abstract consensus model as a probabilistic graphical model, and demonstrated how we query this probabilistic distribution to obtain the  marginal distribution of the joint consensus for each task.
Moreover, we showed how discrete annotation models can be accommodated within our framework, giving pooled Multinomial \cite{Paun2018} and Dawid-Skene \cite{Dawid1979} models as examples. 

Once it fits a probabilistic model to a given annotation data, Crowdnalysis can also leverage this model to conduct a \emph{prospective data quality analysis} for the crowd of interest.
Specifically, Crowdnalysis uses the estimated parameters for tasks and annotators (i.e., marginal probabilities of classes and annotator error rates, respectively) to generate \emph{synthetic} annotation data of different sizes, mimicking crowd behavior. 
Then, performance analysis on this synthetic data provides the expected values of a given metric---say, accuracy---over the ``number of annotations per task''. This analysis allows the determination of the redundancy in annotations needed for the crowd to reach the desired accuracy.
Consequently, given the performances of different communities on the same tasks, prospective analysis also helps choose the community to work with in a future project with similar task characteristics. This feature is particularly beneficial when some communities are not free of charge (paid workers) or hard to assemble (experts).

\begin{figure}[th]
    \begin{subfigure}[t]{.45\textwidth}
        \centering
        \includegraphics[width=\linewidth]{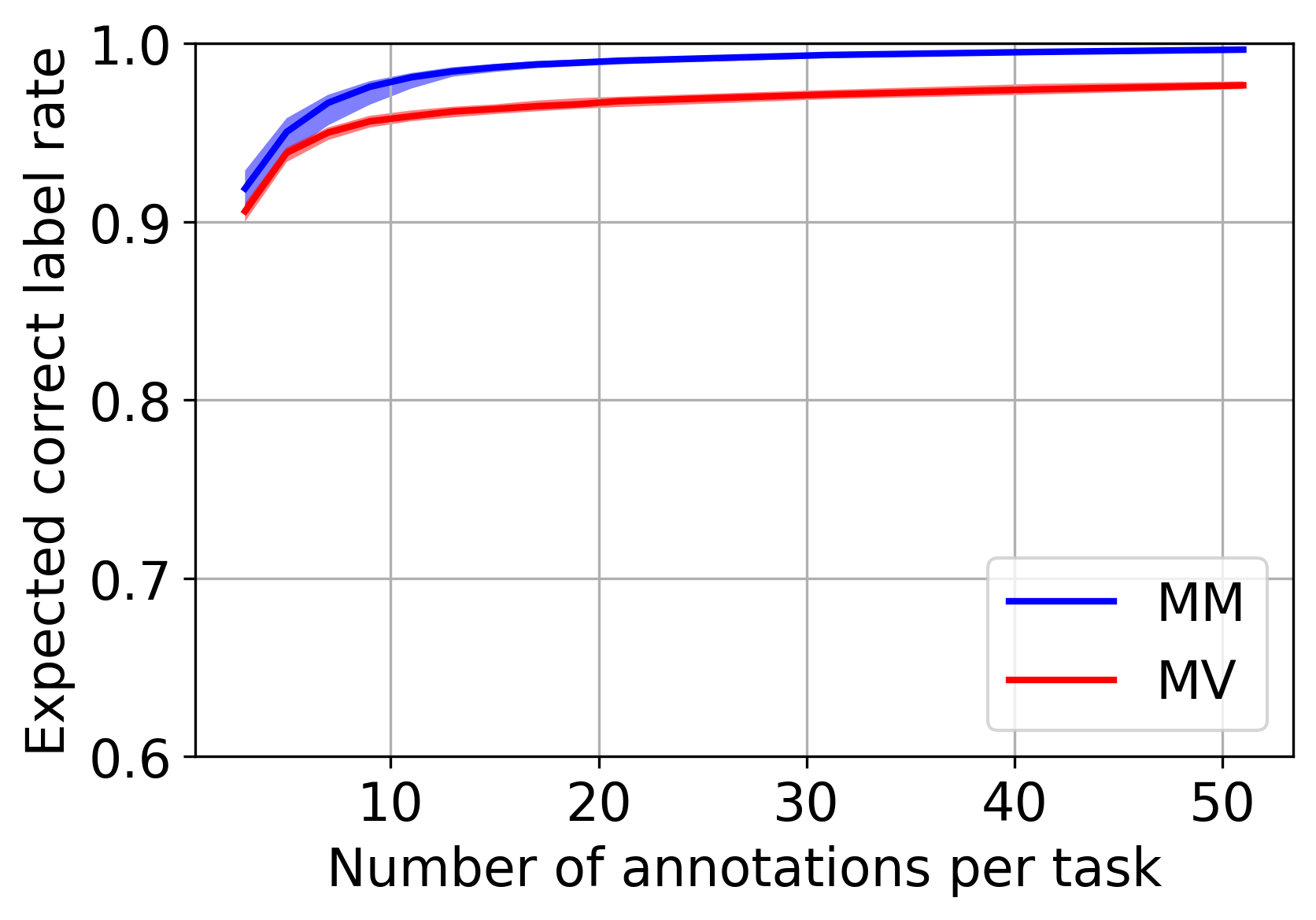}
        \caption{Experts}
        \label{fig:crowdnalysis_accuracy_experts}
    \end{subfigure}
    \hfill
    \begin{subfigure}[t]{.45\textwidth}
        \centering
        \includegraphics[width=\linewidth]{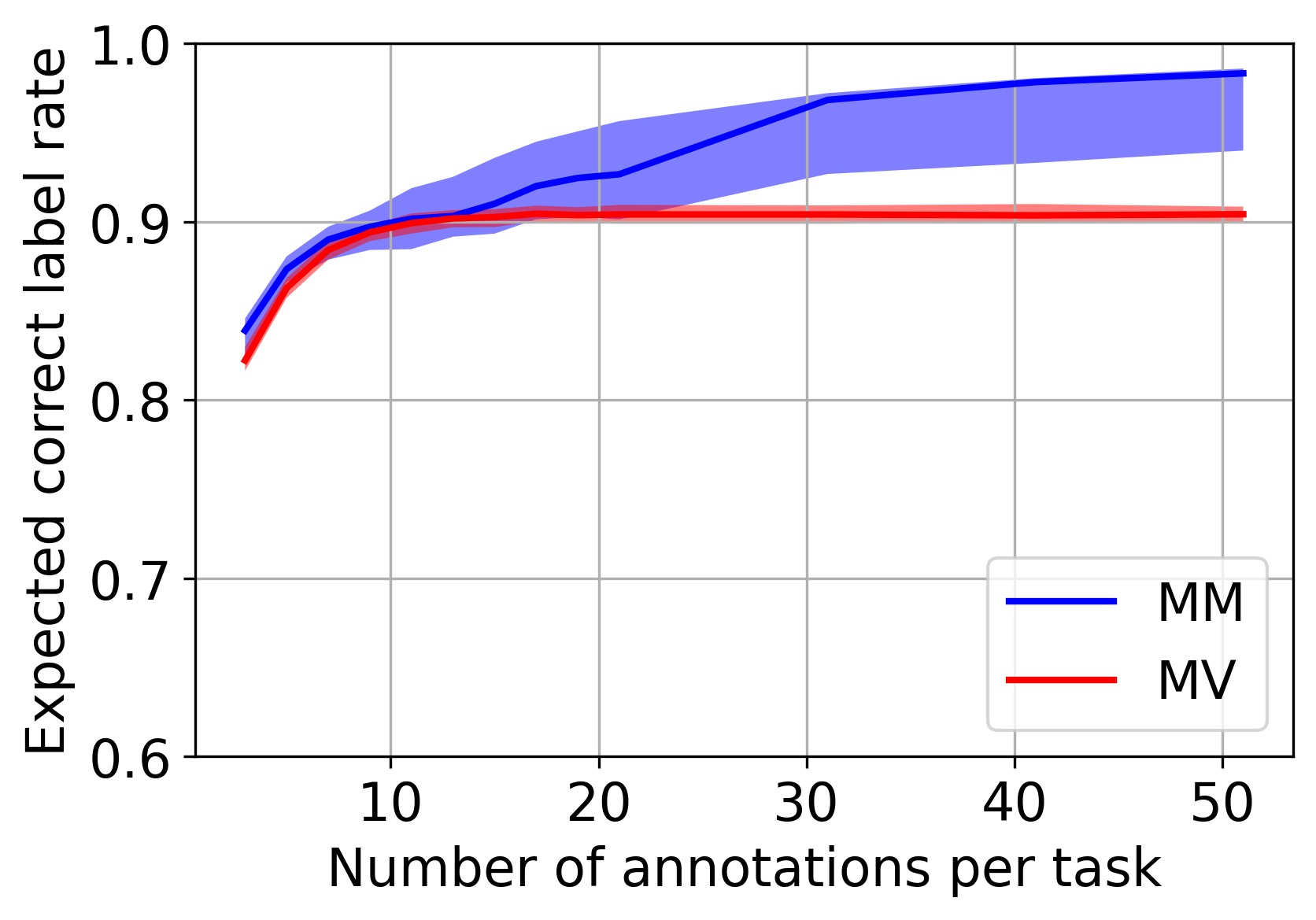}
        \caption{Volunteers}
        \label{fig:crowdnalysis_accuracy_volunteers}
    \end{subfigure}
    \caption{Prospective analysis for correct label rates for experts and volunteers with the Multinomial Model (MM) and Majority Vote (MV) in the 2019 Albania earthquake case study.}
    \label{fig:crowdnalysis_accuracy_all}
\end{figure}

In \cite{cerquides2021}, we present a case study of the 2019 Albania earthquake\footnote{\url{https://en.wikipedia.org/wiki/2019_Albania_earthquake}} analyzing the annotation data from three different communities, namely, volunteers, paid workers and experts, and comparing their performances by Crowdnalysis. All three communities were provided with the same set of social media images and asked to annotate the severity of damages in the photos with one of the labels in $\mathcal{A}=\{$\small\textsf{irrelevant}, \textsf{no-damage}, \textsf{minimal}, \textsf{moderate}, \textsf{severe}\normalsize$\}$.
We provide a concise description of the communities and crowdsourcing settings in Section~\ref{sec:case-studies}. In Fig.~\ref{fig:crowdnalysis_accuracy_all}, we present part of the prospective analysis of this case study. We performed this analysis in three steps: 
1) In the first step, we fit the Multinomial Model (MM) with annotation data from experts and calculated the consensus. Then, we fit another MM to the volunteer data, using the experts' consensus as the ground truth. 
2) In the second step, we used the prior probability distribution for the above labels, estimated by the experts' MM, to generate a dataset of $10,000$ synthetic tasks. Then, we generated synthetic annotation datasets for both experts and volunteers with different redundancies per task using their estimated error rates computed in the first step.
3) In the last step, we computed the consensuses of synthetic expert and volunteer annotations with MM and Majority Vote (MV). Finally, we compared their expected accuracies over the number of annotations per task, using the labels of the synthetic tasks generated in the second step as the ground truth.

In Figs.~\ref{fig:crowdnalysis_accuracy_experts} and~\ref{fig:crowdnalysis_accuracy_volunteers}, we observe that our proposed probabilistic model outperforms the MV. The performance is comparable only for the lowest redundancy, that is , three annotations per task.
The figures also show that when we use the probabilistic model for consensus, the accuracy increases with redundancy in the annotations. 
This observation is also valid for MV up to a certain redundancy value, after which we achieve no, or very minuscule, increase in accuracy.

Owing to Crowdnalysis' probabilistic modeling for consensus and prospective analysis features, CS project managers can make more informed decisions on the crowd to work with and the size of that crowd. 
For example, to obtain a $0.95$ consensus accuracy in a scenario similar to an earthquake, Fig.~\ref{fig:crowdnalysis_accuracy_all} indicates that they would need $6$ expert annotations per task, whereas it would take more than $20$ volunteer annotations for the same accuracy (assuming these two types of annotators exhibit the same or fairly similar annotation skills as the members of the corresponding community that was previously worked with).

Crowdnalysis is implemented with the Python programming language and distributed via the Python Package Index\footnote{\url{https://pypi.org/project/crowdnalysis}} (PyPI) software repository. Therefore, it can be easily imported and used in any Python script. Its source code is publicly available at GitHub\footnote{\url{https://github.com/Crowd4SDG/crowdnalysis}} and bears an exhaustive test suite.
Crowdnalysis also comprises consensus model implementations in the Stan probabilistic programming language~\cite{Stan229}  thanks to the CmdStanPy\footnote{https://mc-stan.org/cmdstanpy} command-line interface for Python. 
Furthermore, Crowdnalysis has the following additional features:
\begin{itemize}
    \item Setting inter-dependencies between questions to filter out irrelevant annotations (e.g., ignoring the following answer for a specific task if the previous answer is ``Not relevant'');
    \item Distinguishing real classes for answers from reported labels (e.g., ``Not answered'');
    \item Calculating inter-rater reliability using different measures (e.g., Fleiss' kappa);
    \item Visualization of annotator error rates and consensus.
\end{itemize}

%\section{Validation}

\section{Validation}
\label{sec:validation}

\subsection{Validation methods}

The quality of the results obtained by crowd-based data analysis depends on different factors: the quality of the  dataset obtained from the previous preparation and analysis phases, the manner in which the task has been managed and described and the skills and expertise of the workers. 
In our case studies we based our validation on the quality of the resulting data. Adopting this end-to-end approach, we can evaluate the quality of the result by considering the effect of other components: the filtering components in VisualCit, which can also present quality issues, as they are based on trained models with machine learning methods, and for which configuration parameters can have an impact on the final results, as well as; the quality of the crowdsourcing tasks assigned to the crowd and the quality of the results of crowdsourcing activities, as described in Section \ref{sec:Crowdnalysis}.

In order to assess such a quality, several methods can be applied:
\begin{itemize}
 \item \emph{Individual}: The role of humans in crowdsourcing can be extended to the quality assessment phase. The accuracy of a given output is evaluated by individuals (e.g., workers or external experts)
 \item \emph{Group}: The assessment is performed by a group of people (typically workers) (e.g., voting).
 \item \emph{Computation-based}: This includes assessment methods that can be performed by a machine without the involvement of humans. 
 \item \emph{Validation datasets}: Validation can be performed by comparing the obtained results with external sources focusing on the same problem (e.g., surveys in the field).
 \end{itemize}

The computation-based methods can support the evaluation of the quality of results from different perspectives and consider different quality dimensions: 
\begin{itemize}
\item \emph{Accuracy}: This can be computed considering a given ground truth and suitable comparison operators to evaluate the similarity between the obtained results and the desired values. The assessment of accuracy can be precise depending on the reference source considered.
\item \emph{Timeliness}: Some data analyses have stringent time constraints. Timeliness measures the temporal validity of the input data; only not outdated values should be considered. This means that the results should be produced within a given time after the event occurs.
\item \emph{Provenance}: The reliability of the input data increases if information about acquisition methods, processing steps, and the way in which crowdsourcing has been performed (e.g., crowd characteristics, redundancy, and aggregation methods) is available. It is important in the evaluation of the results to obtain information on the data provenance: the more input data are reliable, the more trustworthy the analysis results are.
\end{itemize}

%\subsection{Quality control of datasets - UNITAR? 1 page}
A   quality control methodology for datasets derived from social media has been studied in Crowd4SDG to assess whether extracted data can be used as non-traditional data sources by national statistical offices (NSOs) \cite{D5.2}.
A scoring approach was proposed to evaluate the quality of the dataset. In addition to the validation criteria mentioned above, a few new criteria were added for validation focusing on the data production process, confidentiality, and impartiality. 

The results of the validation  for the three case studies are discussed in the next section and listed in Table~\ref{fig:cases}. The main validation criteria are discussed for the results of each of the CSSK components used in the case studies.  The results derived from the quality control methodology are reported in the Data quality assessment column. Finally, a general end-to-end evaluation is given under the General evaluation column and the main issues mentioned under Problems.

\subsection{Case studies}
\label{sec:case-studies}

The following case studies are presented to illustrate the different contexts of data analysis and their challenges and results.
\begin{itemize}
\item \emph{2019 Albania earthquake}:
In this case study, we aimed to provide a working example of how we fit the conceptual and mathematical framework behind Crowdnalysis to real-world data, and demonstrate how we could benefit from our software library in an emergency scenario, where we refer to crowdsourcing. We have published the framework and case study results in \cite{cerquides2021}.

The imagery dataset we worked with was the courtesy of the Qatar Computing Research Institute,  which contains  social media images posted for the 2019 Albanian earthquake filtered by their Artificial Intelligence for Disaster Response (AIDR) platform \cite{imran2014aidr}. The AIDR collected Twitter posts for four days after November 26\textsuperscript{th}, 2019, when the earthquake had occurred, which was the strongest for the country in the last 40 years.
The AIDR automatically classified $907$ images as relevant out of $9,241$ collected.

As mentioned in Section~\ref{sec:Crowdnalysis},  crowdsourcing was performed using three different groups of citizens to annotate the relevant images and the severity of damage seen in the photos by humans compared to the AIDR. Specifically, to obtain a ground truth for the labels, we first worked with ten disaster experts and configured a redundancy of $3$ for annotations on the Crowd4EMS platform \cite{ravi2019crowd4ems}, a precursor of CSPB.
We then referred to a group of 50 volunteers on Crowd4EMS with a redundancy of $3$. Finally, 171 paid workers annotated the same dataset on the Amazon Mechanical Turk (MTurk) platform\footnote{\url{https://www.mturk.com}} with a redundancy of $10$. Using Crowdnalysis, we first computed the experts' consensus as the ground truth and, successively, calculated the error rates---in labeling---of the three crowds. We also carried out a prospective analysis of the communities designed in Section~\ref{sec:Crowdnalysis} of this article.
This case study is related to the use case C in Fig.~\ref{fig:methods}.

Subsequently, we extended our analysis with this dataset incorporating VisualCit for further filtering the AIDR dataset for relevant images; hence, we adopted the use case D in the same figure. 
We provide the confusion matrix for VisualCit in Table~\ref{tab:albania_visualcit}, which shows a $0.83$ precision, a $0.90$ recall and a $0.82$ overall accuracy of VisualCit, with respect to the experts' consensus on relevance.
\vskip -10pt
\begin{table}[!h]
\centering
\begin{tabular}{@{}cc|cc@{}}
\multicolumn{1}{c}{} &\multicolumn{1}{c}{} &\multicolumn{2}{c}{VisualCit} \\
& Relevant & Yes & No \\
\cmidrule{2-4}
\multirow[c]{2}{*}{\rotatebox[origin=c]{90}{Experts}}
& Yes & 546 & 59   \\[1.7ex]
& No & 108 & 194 \\ 
\\
\end{tabular}
\caption{Number of relevant and irrelevant images labeled by VisualCit w.r.t. the experts' consensus.}
\label{tab:albania_visualcit}
\end{table}
\vskip -25pt
Moreover, we conducted a hypothetical experiment to determine whether the results would be accurate if the experts and the MTurk crowd annotated the subset of the AIDR dataset filtered by VisualCit. Specifically, after removing the annotation data for the $253$ tasks excluded by VisualCit, we calculated the new experts' consensus and error rates of the experts and MTurk crowd by Crowdnalysis. We observed that the error rates for both communities remained almost the same (e.g., +$0.01$ in experts' recall, -$0.04$ in MTurk crowd's specificity).

Our extended analysis suggests that when we can trade crowdsourcing efforts for higher recall, we can safely bring VisualCit to automatically filter relevant images, thus; reducing the number of irrelevant tasks to be processed by crowd workers.

\item \emph{COVID-19 social distancing case study}: This case study, developed in 2020 during the initial phases of the pandemic, aimed to derive indicators of social distancing behavior and face mask usage in different countries
\cite{DBLP:conf/icse/NegriSARSSFCP21}.
This case study is related to use case 
F shown in Fig.~\ref{fig:methods}. Selected raw data were captured from social media from May to August 2020, crawling Twitter with generic COVID-related keywords. Filtering posts with VisualCit, selecting only images that are classified as photos (excluding memes, maps, drawings, etc.), in outdoor spaces, and containing at least two persons. When a native geolocation was not present in the tweet, the posts were geolocated using the CIME algorithm \cite{scalia2021cime}. Crowdsourcing performed with the CS Project Builder was used to confirm the relevance  according to the above-mentioned criteria and the automatic geolocation of the images, and to assess whether face masks were correctly used and social distancing rules followed. 
The majority vote was used to compute the consensus on the crowd annotations. The resulting images were aggregated to build behavioral indicators on a national basis.

%flood and gender \cite{genderandfloods21}

%Nepal and 
\item \emph{Thailand flood alerting case study}: In this case study, a large-scale flood event in Thailand was evaluated. Tropical storm Dianmu\footnote{\url{https://floodlist.com/asia/thailand-tropical-storm-dianmu-floods-september-2021}} hit Thailand in late September and October 2021. According to the UNOSAT Thailand flood monitoring dashboard\footnote{\url{https://unosat-geodrr.cern.ch/portal/apps/opsdashboard/index.html#/4f878691713a40f3b8ef3140e63c9f6d}}, as a preliminary assessment around 1.4 million people were affected by flooding. The goal of this case study was to assess the ability to capture the onset of large scale emergencies, while enabling a spatial description of the event. This case study refers to the use of Cases G and H in Fig. \ref{fig:methods}. An overall description of this approach can be found in \cite{bono2022triggercit}. Tweets corresponding to event onset were crawled using a small set of flood-related Thai keywords, provided by the UNOSAT office in Bangkok. The time series corresponding to keywords' usage during event onset is shown in Fig. \ref{fig:thai_counts_2}. After detecting the onset trend, a VisualCit pipeline was executed to remove near-duplicates, non-photos, and NSFW images, to exclude irrelevant content. The filtered posts were then geolocated with CIME, focusing on locations within Thailand's national borders. The results were then aggregated by administrative region and normalized using the population count.

\begin{figure}[th!]
    \centering
    \includegraphics[width=\textwidth]{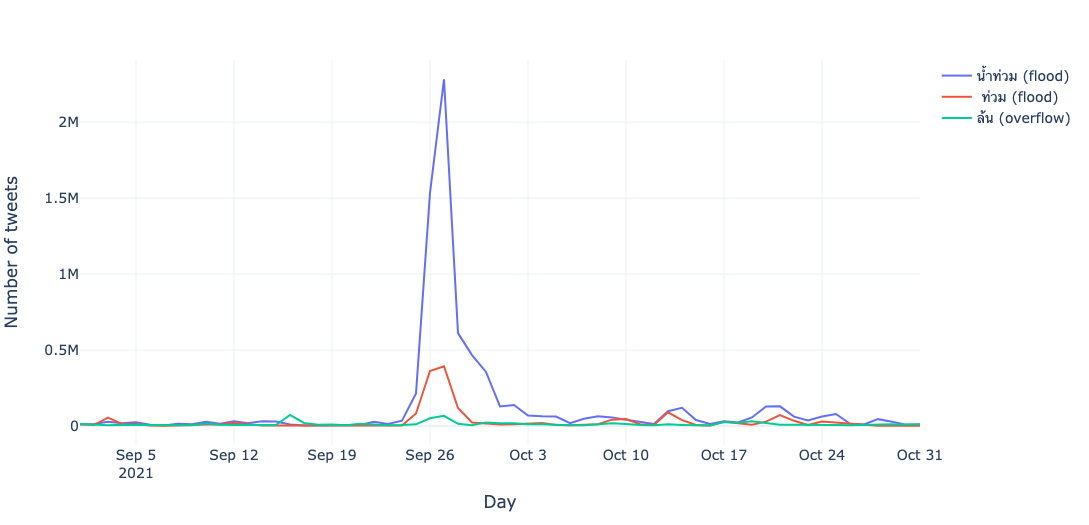}
    \caption{Daily tweet counts for Thai dictionary entries.}
    \label{fig:thai_counts_2}
\end{figure}

\begin{comment}

\begin{figure}[th!]
    \centering
    \subfigure[Native Twitter geolocation]{\includegraphics[width=0.245\textwidth]{figures/visualcit/thailand_pos_native.png}}   
    \subfigure[VisualCit geolocation]{\includegraphics[width=0.245\textwidth]{figures/visualcit/thailand_pos_cime.png}} 
    \subfigure[VisualCit geolocation with extended tweet set]{\includegraphics[width=0.45\textwidth]{figures/visualcit/thailand_pos_come_promising.png}}
    \subfigure[Affected people]{\includegraphics[width=0.245\textwidth]{figures/visualcit/thailand_affected.png}}
    \caption{Geolocations / inhabitants ratio by region (a) Twitter native geolocations, (b) Twitter native + CIME geolocated, (c) Twitter native + CIME geolocated from extended dataset (with images + promising text-only tweets), and (d) Number of affected persons by region at September, 28th (source: ReliefWeb).}
    \label{fig:thaigeo}
\end{figure}

\end{comment}

\end{itemize}

\begin{table*}[th]
    \centering
    \includegraphics[width=\textwidth]{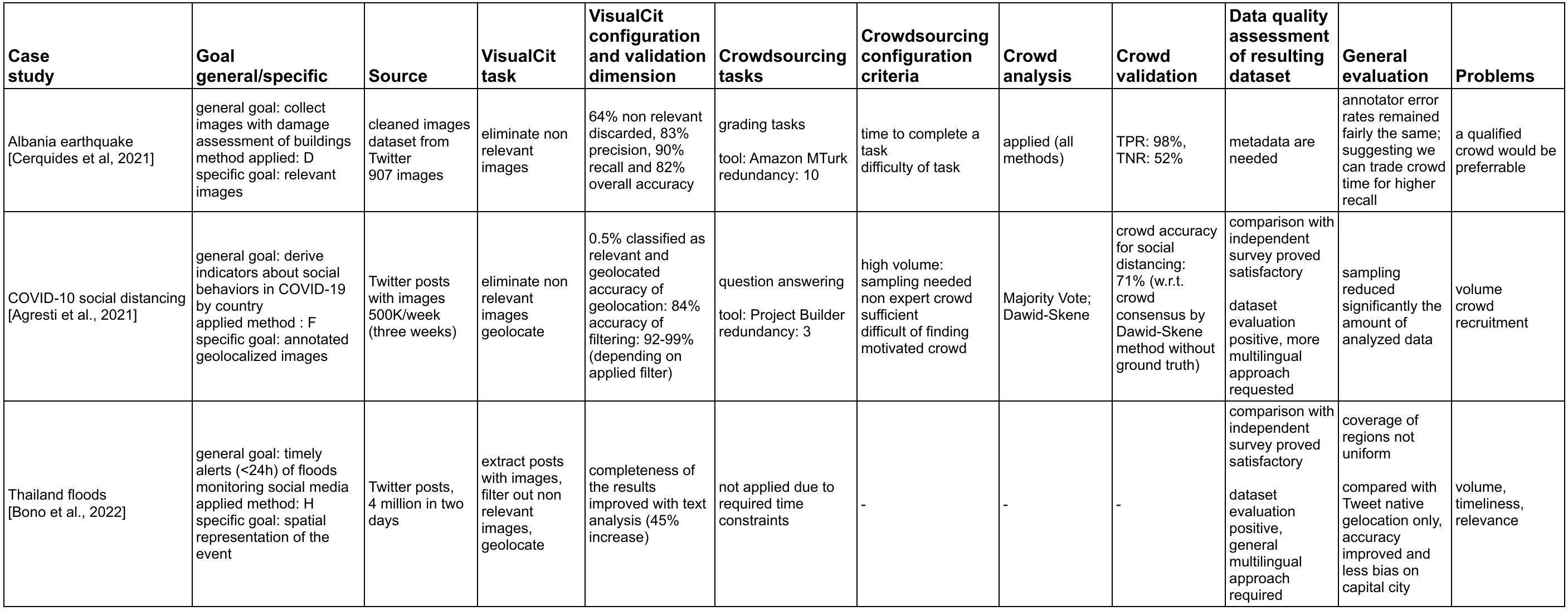}
    \caption{Case studies.}
    \label{fig:cases}
\end{table*}

The aggregate descriptions of the results and insights from the case studies are represented in Table~\ref{fig:cases}.

We present here the main results and outcomes of the case studies presented above and, discuss their validation procedure.

These three case studies were aimed at a range of goals. In the Albania case study, the focus was on analyzing images from posts assessing damage to buildings, while in the COVID and Thailand case studies, the goal was to generate thematic maps providing evidence about ongoing situations.

The case studies were characterized using different types of source data. In most cases, social media data are obtained through the crawling of Twitter, while in one case (the Albania case study), the initial dataset is available as the result of previous analyses. Consequently, the volume of posts to be analyzed was significantly different. In addition to the volume aspect, the three case  studies were characterized by different velocity characteristics, and timeliness requirements. In particular, the Thailand flood case study has a large number of posts to be analyzed rapidly to provide a rapid assessment of the situation, while in the COVID case study, the volume is high, while timeliness is less stringent, as the analysis is performed on a weekly basis.
The VisualCit results are validated in different ways, according to the tasks to be performed. In the Albania case study, VisualCit was used to further pinpoint relevant images from the initial dataset, and a significant improvement in accuracy was obtained.
In the COVID case study, filtering resulted in only 0.5\% of the images being considered relevant for the task to be performed. A detailed evaluation of precision and recall for different filters was published in \cite{DBLP:conf/icse/NegriSARSSFCP21}, and the accuracy of geolocation measured on a country basis was 84\%. In the Thailand case study, low recall was compensated by analyzing the text of the tweets, thus improving completeness in the data for the considered regions.
Crowdsourcing was only applied to the first two case studies. In Albania case study, crowdsourcing was performed using three different communities: a group of experts, a group of volunteers, and a group of paid workers on Amazon Turk, as detailed above. In the COVID case study, characterized by a high volume even notwithstanding the high VisualCit selectivity, the CS Project Builder was used with redundancy 3 and majority vote. The main problem encountered in this case was related to the size of the data and the availability of a crowd of sufficient size for performing the analysis, and the costs of the analysis when a paid crowd is employed.
In the Albania case study, the performance of the crowd was analyzed using Crowdnalysis, based on comparisons with expert evaluations,  as illustrated in Section~\ref{sec:Crowdnalysis}. However, the size of the COVID case study did not allow extensive data validation of the results, so an indirect validation procedure was adopted, comparing the results with similar analyses conducted at the same time with surveys, highlighting a significant correlation of the results \cite{DBLP:conf/icse/NegriSARSSFCP21}.
In addition, in the case of the Thailand flood case study, validation was performed by comparing the obtained data with data collected in the field, showing good qualitative results of the method \cite{bono2022triggercit}.

The selected case studies represent various goals and conditions. 
%The variety of the case studies resulted in general results and problems to be studied. 
In general, the volume of data is high, and timeliness may be a constraint, whereas the number of relevant tweets for a given goal is usually small. Fast automatic filtering tools, such as VisualCit can be successfully employed to speed up the analysis and reduce the number of irrelevant elements to be examined. When timeliness poses constraints of less than 24h to analyze a situation, acceptable results can be obtained with completely automated tools, while better accuracy can be obtained by adding crowdsourcing if time is available for the analysis.
In general, while citizen scientists for crowdsourcing activities are difficult or costly to recruit, better results can be obtained with a selected crowd or experts, thus showing the importance of crowd analysis.

\section{Related work}
\label{sec:sota}

%\subsection{Emergencies and social media analysis}  
Social media can provide timely information about ongoing events \cite{social-emergency22, DBLP:journals/csur/ImranCDV15}. Their potential has been studied in several contexts, such as earthquakes \cite{DBLP:journals/tkde/SakakiOM13}, demonstrating their significant timeliness that can help prevent further losses.

Social media analysis, including both an automatic analysis of images contained in posts and the collaboration of human computing, has been discussed by several authors in the literature, for example, 
\cite{havas2017e2mc, ravi2019crowd4ems, DBLP:journals/ijhci/AlamOI18}.
For automatic analysis of images, many recent approaches are based on AI and in particular neural networks, for example, \cite{IMRAN2020102261,pennington2022near,ofli2022real}. Hybrid deep-learning and crowdsourcing approaches were analyzed in \cite{anjum2021exploring}.
The opportunities and challenges of using social media in emergencies were analyzed in
\cite{wiegmann2021opportunities}.
Several challenges are posed by the need of geolocating posts, in order to be able to leverage their contents.
%\subsection{Social media analysis - Barbara and Mark}
Several approaches have been proposed in the literature, for example, \cite{DBLP:journals/expert/MiddletonMM14}, and the issues of analyzing posts in multilingual contexts are being studied \cite{scalia2021cime,lorini2019integrating}. While the precision of the location is important, it is difficult to achieve, and it can not rely on native locations of tweets, both due to their scarcity, but also because the location of the posting is often different from the location of the  event during emergencies.
In this study, we focused on combining automatic analysis of posts and their images and their geolocation and crowdsourcing as an approach to refine this automatic analysis gather further information, and support the selection of relevant posts, with a human-in-the-loop approach in all phases.

%\subsection{Crowdsourcing and citizen science}
In \cite{lukyanenko2020citizen}, some successful projects based on \CS~were discussed, showing their potential to  analyze and solve problems where the scientific community is not investigating a specific problem or resources are scarce.

\CS~has been advocated to support the collection of information for indicators of Sustainable Development Goals (SDG). In the paper ``Goals Mapping \CS~contributions to the UN sustainable development goals'' \cite{fritz2019citizen}, the authors perform a detailed analysis of existing projects covering some of the SDG indicators and also potential areas for further development of new initiatives, where collection of information is difficult from other sources; in particular, the potential for climate change indicators is mentioned. 
Crowdsourcing (or crowd science) and \CS~are often studied in separate research communities. The potential of combining their approaches in a multi-dimensional framework was discussed in \cite{franzoni2022crowds}.
Within Crowd4SDG, the potential of social media was discussed in \cite{D5.1}, considering lessons learned from case studies for SDG monitoring, to provide key statistics and discuss also the role of local indicators.

\begin{comment}
crowdsourcing tools ({\bf Mark for crowdsourcing})

consensus mechanisms (+ Mark and revision by Cinzia)
\end{comment}

The primary goal of every crowdsourcing activity is to obtain reliable ground truth labels for a set of tasks of interest. The ground truth for each task is calculated by aggregating the crowd members' answers to the question asked for that task.
The reliability of the consensus depends on several factors, including the relevance of the tasks to the problem at hand, the accuracy and expertise of the annotators, the redundancy in answers, and the soundness of the protocol asked to follow for annotation.

The intrinsic complexity of the label aggregation problem has received considerable attention from the statistics and machine learning communities.
Researchers have introduced different measures to assess the extent of agreement among annotators (i.e, raters) \cite{scott1955, fleiss1971}. Gwet gathered a wide collection of inter-rater reliability measures, including his own, in~\cite{Gwet2014}.
Employing such a measure on crowdsourced data would be a good practice before calculating the ground truth itself, as the annotations of a poorly agreeing crowd would probably not yield a trustworthy result. However, a high agreement rate does not imply high accuracy when the annotators agree on the wrong label. Therefore, modeling individual annotator behavior may prove essential depending on the crowd profile.

Accordingly, although the majority vote is still a common choice for consensus calculation, many researchers have used probabilistic consensus models to account for the uncertainties and deficiencies in annotation data and the variation in annotator skills.
The seminal work of Dawid and Skene~\cite{Dawid1979} established a benchmark algorithm that can iteratively estimate both the individual annotator error rates and consensus on tasks when the ground truth is unknown. A simpler approach that does not differentiate annotators is the multinomial model proposed in~\cite{Paun2018}. Passonneau and Carpenter~\cite{Passonneau2014} discussed the benefits of relying on annotation models and demonstrated how these models can reduce the cost of obtaining crowdsourced data and provide confidence for the ground truth.
Another probabilistic model, GLAD~\cite{Whitehill2009}, also estimates the difficulty of each task, which can be leveraged in task assignments by means of active sampling.
Paun et al.~\cite{Paun2018} provided a comparison of Bayesian annotation models by categorizing the models as \textit{pooled} (all annotators had the same ability), \textit{unpooled} (individual models of annotators) and \textit{partially pooled} (hierarchical model of abilities). Cerquides~\cite{Cerquides2021b} introduced a new understanding of hierarchical models using geometric intuitions.
CrowdTruth framework~\cite{Dumitrache2018a} takes a rather methodological approach by modeling three components of a crowdsourcing system, namely, ``input media units, workers, and annotations''. This framework has been inspirational for the conceptual framework of Crowdnalysis~\cite{cerquides2021}, as detailed in Section~\ref{sec:Crowdnalysis}. CrowdTruth draws inspiration from Aroyo and Welty's work~\cite{Aroyo2015}, where the authors present seven misconceptions in crowdsourcing, such as there can be only ``one truth'' for an input task and ``experts are better'' than non-experts.
Furthermore, Li presented a theoretical analysis of crowdsourcing algorithms ~\cite{Li2015}.
\section{Concluding remarks}
\label{sec:conclusions}
In this paper, we presented an innovative approach to support CS project managers and analysts in the analysis of social media data using a combination of automatic classification, filtering, and geolocation tools and a human-in-the-loop \CS~approach. The proposed “data analysis pipeline” retrieves information from large sets of posts---many of which may be irrelevant---to configure automated AI-based filtering tools, to create crowdsourcing projects to gather inputs from citizens. We also illustrate the structure and use of dedicated tools to support the CS project implementation in each step of the process. These tools were developed within the Crowd4SDG European project and are  included in the ``Citizen Science Solution Kit'' (CSSK). A systematic validation of the approach with three case studies is presented, and the validation criteria are discussed. Further details on the tools can be found on the project web site\footnote{\url{https://crowd4sdg.eu/about-2/tools/}}.

The research community is currently very active in developing tools for automatically analyzing social media. However, collecting such information and images may introduce several types of bias that should still be investigated. These include biases in the analysis due to variations in the profiles and densities of contributors in different areas, that is, there is a risk of underrepresenting minorities and focusing only on a subset of events related to the profiles of the contributors. In addition, analyzing social media in isolation presents all the limitations of a single source, and efforts should be made to integrate other sources of information. Systematic sampling, data augmentation, and bias reduction techniques should be investigated. Near real-time integration with other sources of information is also to be exploited, such as different social media, Voluntary Geographical Information such as OpenStreetMap \cite{DBLP:journals/pervasive/HaklayW08}, Earth observations such as Google Earth or sensor networks, official reporting agencies, and statistical information. The challenges of  this integration on a global scale are high, because of the variety of available data; however, tools such as OpenStreetMap, which provide a rich source of crowdsourced information beyond geographical locations, if integrated into social media analysis, would greatly increase the understanding of events. 
Future work also includes incorporating hierarchical consensus models into Crowdnalysis and leveraging annotator models to efficiently govern task-annotator assignments within CSPB to make the most out of citizen contributions. Given that local knowledge can be favorable in crowdsourcing and emergency responses, we also plan to leverage the geolocation information to assign tasks to best-performing annotators in related regions, if available.

The development of tools has been a major goal for the Crowd4SDG, based on ongoing developments in partner institutions. The CSSK responds to the needs of researchers approaching crowdsourcing methodology for the first time, providing them with easy-to-use and flexible tools for testing and implementing their
projects. This is particularly important when aiming to support National Statistical Offices in the collection and analysis of data for official SDG reporting, as highlighted in the Crowd4SDG policy brief on using \CS~data to track SDG progress \cite{policybrief}. 

Several academic institutions and Non-Governmental Organizations have used the Citizen Science Project Builder tool in CSSK to setup projects in various scientific disciplines and social fields. Owing to the develop and enhancements of the platform carried out in the context of the Crowd4SDG partnership, CSSK has been used specifically to test different approaches to crowdsourced disaster response, as described in the case studies and examined in the Crowd4SDG challenges. Multiple organizations have already profited from the more performing interface and the smoother user experience, including researchers at the Joint Research Centre of the European Commission. However, extensive outreach activities are needed to make the wider emergency community aware of the availability of the CSSK, and  knowledgeable in the use of its different components.

Social media activities as a form of “passive” CS contribution, where useful information can be discovered through crawling and analysis, represent huge and untapped data. 
On the other hand, passive contributions complement---equally, if not more powerful---“active” ways for citizens to contribute to data collection and analysis.
%Passive contributions however complement several others, equally if not more powerful, “active” ways for citizens to contribute both data and data analysis tasks. 
When active participation of citizens is required, for example by some of the CS tools, crowdsourcing poses several known challenges related to engaging and retaining participants. These include making sure that citizens and their communities benefit from participation, planning adequate rewards to recruit and keep the crowd active and interested, and the necessity to tailor the tasks to the variety of possible skills. However, solutions to these issues requires further experimentation and development.

\section*{Acknowledgements}
This work was  funded by the European Commission H2020 project  Crowd4SDG  ``Citizen Science for Monitoring Climate Impacts and Achieving Climate Resilience'', \#872944. 
This work expresses the opinions of the authors and not necessarily those of the European Commission. The European Commission is not liable for any use that may be made of the information contained in this work. 

The authors thank Miguel Armendáriz Jáuregui for his work on the VisualCit interface and all the students who experimented with CSSK during its development.

The authors thank all the Crowd4SDG partners for the discussions and common work during the project.

\bibliographystyle{plain}
\bibliography{biblio.bib}

\begin{thebibliography}{10}

\bibitem{social-emergency22}
Anouck Adrot, Samuel Auclair, Julien Coche, Audrey Fertier, Cécile Gracianne,
  and Aurélie Montarnal.
\newblock Using social media data in emergency management: a proposal for a
  socio-technical framework and a systematic literature review.
\newblock In {\em Proc. ISCRAM, 2022}, May 2022.

\bibitem{DBLP:journals/ijhci/AlamOI18}
Firoj Alam, Ferda Ofli, and Muhammad Imran.
\newblock Processing social media images by combining human and machine
  computing during crises.
\newblock {\em Int. J. Hum. Comput. Interaction}, 34(4):311--327, 2018.

\bibitem{anjum2021exploring}
Samreen Anjum, Ambika Verma, Brandon Dang, and Danna Gurari.
\newblock Exploring the use of deep learning with crowdsourcing to annotate
  images.
\newblock {\em Human Computation}, 8(2):76--106, 2021.

\bibitem{Aroyo2015}
Lora Aroyo and Chris Welty.
\newblock {Truth Is a Lie: Crowd Truth and the Seven Myths of Human
  Annotation}.
\newblock {\em AI Magazine}, 36(1):15--24, 2015.

\bibitem{asif2021automatic}
Amna Asif, Shaheen Khatoon, Md~Maruf Hasan, Majed~A Alshamari, Sherif Abdou,
  Khaled~Mostafa Elsayed, and Mohsen Rashwan.
\newblock Automatic analysis of social media images to identify disaster type
  and infer appropriate emergency response.
\newblock {\em Journal of Big Data}, 8(1):1--28, 2021.

\bibitem{bono2022triggercit}
Carlo Bono, Barbara Pernici, Jose~Luis Fernandez-Marquez, Amudha~Ravi Shankar,
  Mehmet~O{\u{g}}uz M{\"u}l{\^a}yim, and Edoardo Nemni.
\newblock {TriggerCit: Early Flood Alerting using {Twitter} and Geolocation--a
  comparison with alternative sources}.
\newblock In {\em Proceedings of ISCRAM 2022, Tarbes, France}, May 2022.

\bibitem{Cerquides2021b}
Jesus Cerquides.
\newblock A first approach to closeness distributions.
\newblock {\em Mathematics}, 9(23):3112, Dec 2021.

\bibitem{crowdnalysis2022}
Jesus Cerquides and Mehmet~O{\u{g}}uz M{\"{u}}l{\^{a}}yim.
\newblock crowdnalysis: A software library to help analyze crowdsourcing
  results.
\newblock https://doi.org/10.5281/zenodo.5898579, January 2022.

\bibitem{cerquides2021}
Jesus Cerquides, Mehmet~Oğuz Mülâyim, Jerónimo Hernández-González, Amudha
  Ravi~Shankar, and Jose~Luis Fernandez-Marquez.
\newblock A conceptual probabilistic framework for annotation aggregation of
  citizen science data.
\newblock {\em Mathematics}, 9(8):875, 2021.

\bibitem{data-democracy}
Max Craglia and Lea Shanley.
\newblock Data democracy–increased supply of geospatial information and
  expanded participatory processes in the production of data.
\newblock {\em International Journal of Digital Earth}, 8(9):679--693, 2015.

\bibitem{Dawid1979}
Alexander~Philip Dawid and Allan~M. Skene.
\newblock {Maximum Likelihood Estimation of Observer Error-Rates Using the EM
  Algorithm}.
\newblock {\em Applied Statistics}, 28(1):20, 1979.

\bibitem{Dumitrache2018a}
Anca Dumitrache, Oana Inel, Lora Aroyo, Benjamin Timmermans, and Chris Welty.
\newblock {CrowdTruth 2.0: Quality Metrics for Crowdsourcing with
  Disagreement}, aug 2018.

\bibitem{fleiss1971}
Joseph~L Fleiss.
\newblock Measuring nominal scale agreement among many raters.
\newblock {\em Psychological bulletin}, 76(5):378, 1971.

\bibitem{fohringer2015social}
Joachim Fohringer, Doris Dransch, Heidi Kreibich, and Kai Schr{\"o}ter.
\newblock Social media as an information source for rapid flood inundation
  mapping.
\newblock {\em Natural Hazards and Earth System Sciences}, 15(12):2725--2738,
  2015.

\bibitem{franzoni2022crowds}
Chiara Franzoni, Marion Poetz, and Henry Sauermann.
\newblock Crowds, citizens, and science: a multi-dimensional framework and
  agenda for future research.
\newblock {\em Industry and Innovation}, 29(2):251--284, 2022.

\bibitem{fritz2019citizen}
Steffen Fritz, Linda See, Tyler Carlson, Mordechai~Muki Haklay, Jessie~L
  Oliver, Dilek Fraisl, Rosy Mondardini, Martin Brocklehurst, Lea~A Shanley,
  Sven Schade, et~al.
\newblock Citizen science and the united nations sustainable development goals.
\newblock {\em Nature Sustainability}, 2(10):922--930, 2019.

\bibitem{grasso2016codified}
Valentina Grasso and Alfonso Crisci.
\newblock Codified hashtags for weather warning on twitter: an italian case
  study.
\newblock {\em PLoS currents}, 8, 2016.

\bibitem{Grey2009}
Francois Grey.
\newblock The age of citizen cyberscience.
\newblock http://cerncourier.com/cws/article/cern/38718. Accessed Jul 2022,
  2009.

\bibitem{Gwet2014}
Kilem~L Gwet.
\newblock {\em {Handbook of Inter-Rater Reliability: The Definitive Guide to
  Measuring the Extent of Agreement Among Raters}}.
\newblock Advanced Analytics, LLC, 4 edition, 2014.

\bibitem{DBLP:journals/pervasive/HaklayW08}
Mordechai~(Muki) Haklay and Patrick Weber.
\newblock {OpenStreetMap}: {U}ser-generated street maps.
\newblock {\em {IEEE} Pervasive Computing}, 7(4):12--18, 2008.

\bibitem{havas2017e2mc}
Clemens Havas, Bernd Resch, Chiara Francalanci, Barbara Pernici, Gabriele
  Scalia, Jose~Luis Fernandez-Marquez, Tim Van~Achte, Gunter Zeug, Maria
  Rosa~Rosy Mondardini, Domenico Grandoni, et~al.
\newblock E2mc: Improving emergency management service practice through social
  media and crowdsourcing analysis in near real time.
\newblock {\em Sensors}, 17(12):2766, 2017.

\bibitem{DBLP:journals/csur/ImranCDV15}
Muhammad Imran, Carlos Castillo, Fernando Diaz, and Sarah Vieweg.
\newblock Processing social media messages in mass emergency: {A} survey.
\newblock {\em {ACM} Comput. Surv.}, 47(4):67:1--67:38, 2015.

\bibitem{imran2014aidr}
Muhammad Imran, Carlos Castillo, Ji~Lucas, Patrick Meier, and Sarah Vieweg.
\newblock {AIDR: Artificial intelligence for disaster response}.
\newblock In {\em Proceedings of the 23rd International Conference on World
  Wide Web}, pages 159--162, 2014.

\bibitem{IMRAN2020102261}
Muhammad Imran, Ferda Ofli, Doina Caragea, and Antonio Torralba.
\newblock {Using AI and Social Media Multimodal Content for Disaster Response
  and Management: Opportunities, Challenges, and Future Directions}.
\newblock {\em Information Processing \& Management}, 57(5):102261, 2020.

\bibitem{special-track}
Hedi Karray, Antonio~De Nicola, Nada Matta, and Hemant Purohit, editors.
\newblock {\em ISCRAM 2022 Conference Proceedings - 19th International
  Conference on Information Systems for Crisis Response and Management}.
\newblock ISCRAM Digital Library, 2022.

\bibitem{Li2015}
Hongwei Li.
\newblock {\em {Theoretical Analysis and Efficient Algorithms for
  Crowdsourcing}}.
\newblock PhD thesis, UC Berkeley, 2015.

\bibitem{lorini2019integrating}
Valerio Lorini, Carlos Castillo, Francesco Dottori, Milan Kalas, Domenico
  Nappo, and Peter Salamon.
\newblock Integrating social media into a pan-european flood awareness system:
  a multilingual approach.
\newblock {\em arXiv preprint arXiv:1904.10876}, 2019.

\bibitem{lukyanenko2020citizen}
Roman Lukyanenko, Andrea Wiggins, and Holly~K Rosser.
\newblock Citizen science: An information quality research frontier.
\newblock {\em Information Systems Frontiers}, 22(4):961--983, 2020.

\bibitem{DBLP:journals/expert/MiddletonMM14}
Stuart~E. Middleton, Lee Middleton, and Stefano Modafferi.
\newblock Real-time crisis mapping of natural disasters using social media.
\newblock {\em {IEEE} Intelligent Systems}, 29(2):9--17, 2014.

\bibitem{DBLP:conf/icse/NegriSARSSFCP21}
Virginia Negri, Dario Scuratti, Stefano Agresti, Donya Rooein, Gabriele Scalia,
  Amudha~Ravi Shankar, Jose~Luis Fernandez{-}Marquez, Mark~James Carman, and
  Barbara Pernici.
\newblock Image-based social sensing: Combining {AI} and the crowd to mine
  policy-adherence indicators from twitter.
\newblock In {\em 43rd {IEEE/ACM} International Conference on Software
  Engineering: Software Engineering in Society, {ICSE} {(SEIS)} 2021, Madrid,
  Spain, May 25-28, 2021}, pages 92--101. {IEEE}, 2021.

\bibitem{nguyen2017automatic}
Dat~Tien Nguyen, Firoj Alam, Ferda Ofli, and Muhammad Imran.
\newblock Automatic image filtering on social networks using deep learning and
  perceptual hashing during crises.
\newblock In {\em Proc. 14th ISCRAM Conf. – Albi, France}, 5 2017.

\bibitem{ofli2022real}
Ferda Ofli, Umair Qazi, Muhammad Imran, Julien Roch, Catherine Pennington,
  Vanessa Banks, and Remy Bossu.
\newblock A real-time system for detecting landslide reports on social media
  using artificial intelligence.
\newblock {\em arXiv preprint arXiv:2202.07475}, 2022.

\bibitem{Passonneau2014}
Rebecca~J. Passonneau and Bob Carpenter.
\newblock {The Benefits of a Model of Annotation}.
\newblock {\em Transactions of the Association for Computational Linguistics},
  2:311--326, dec 2014.

\bibitem{Paun2018}
Silviu Paun, Bob Carpenter, Jon Chamberlain, Dirk Hovy, Udo Kruschwitz, and
  Massimo Poesio.
\newblock {Comparing Bayesian Models of Annotation}.
\newblock {\em Transactions of the Association for Computational Linguistics},
  6:571--585, dec 2018.

\bibitem{pennington2022near}
Catherine~VL Pennington, R{\'e}my Bossu, Ferda Ofli, Muhammad Imran, Umair~W
  Qazi, Julien Roch, and Vanessa~J Banks.
\newblock A near-real-time global landslide incident reporting tool
  demonstrator using social media and artificial intelligence.
\newblock {\em International Journal of Disaster Risk Reduction}, page 103089,
  2022.

\bibitem{D5.1}
Elena Proden.
\newblock {Crowd4SDG Deliverable 5.1 - Initial report on relevance and
  quality-related considerations of citizen-science generated data}.
\newblock
  https://crowd4sdg.eu/wp-content/uploads/2021/10/D5.1-Initial-report-on-relevance-and-quality.pdf.

\bibitem{D5.2}
Elena Proden.
\newblock {Crowd4SDG Deliverable 5.2 - Data usability assessment and
  recommendations for SDGs GEAR cycle 1}.
\newblock
  https://crowd4sdg.eu/wp-content/uploads/2021/10/D5.2-Data-usability-assessment.pdf,
  2021.

\bibitem{policybrief}
Elena Proden, Karen Bett, Haoyi Chen, Sara~Duerto Valero, Dilek Fraisl, Gabriel
  Gamez, Stephen MacFeely, Rosy Mondardini, Linda See, and Yongyi Min.
\newblock {Citizen science data to track SDG progress: Low-hanging fruit for
  Governments and National Statistical Offices}.
\newblock \url{https://tinyurl.com/crowd4sdg-policy-brief}, 2022.

\bibitem{ravi2019crowd4ems}
Amudha Ravi~Shankar, Jose~Luis Fernandez-Marquez, Barbara Pernici, Gabriele
  Scalia, Maria~Rosa Mondardini, and Giovanna Di~Marzo~Serugendo.
\newblock {Crowd4Ems}: A crowdsourcing platform for gathering and geolocating
  social media content in disaster response.
\newblock {\em International Archives of the Photogrammetry, Remote Sensing and
  Spatial Information Sciences}, 42:331--340, 2019.

\bibitem{DBLP:journals/tkde/SakakiOM13}
Takeshi Sakaki, Makoto Okazaki, and Yutaka Matsuo.
\newblock Tweet analysis for real-time event detection and earthquake reporting
  system development.
\newblock {\em {IEEE} Trans. Knowl. Data Eng.}, 25(4):919--931, 2013.

\bibitem{scalia2021cime}
Gabriele Scalia, Chiara Francalanci, and Barbara Pernici.
\newblock Cime: Context-aware geolocation of emergency-related posts.
\newblock {\em GeoInformatica}, 26:125–--157, 2022.

\bibitem{scott1955}
William~A Scott.
\newblock Reliability of content analysis: The case of nominal scale coding.
\newblock {\em Public Opinion Quarterly}, 19(3):321--325, 1955.

\bibitem{scotti2020enhanced}
Vincenzo Scotti, Mario Giannini, and Francesco Cioffi.
\newblock Enhanced flood mapping using synthetic aperture radar (sar) images,
  hydraulic modelling, and social media: A case study of hurricane harvey
  (houston, tx).
\newblock {\em Journal of Flood Risk Management}, 13(4):e12647, 2020.

\bibitem{Stan229}
{Stan Development Team}.
\newblock {Stan Modeling Language Users Guide and Reference Manual}.
\newblock \url{https://mc-stan.org}, 2022.

\bibitem{Strasser2018}
Bruno Strasser, Jérôme Baudry, Dana Mahr, Gabriela Sanchez, and Elise
  Tancoigne.
\newblock Citizen science? rethinking science and public participation.
\newblock {\em Science \& Technology Studies}, 32(2):52--76, 2019.

\bibitem{wald2016design}
Dara~M Wald, Justin Longo, and AR~Dobell.
\newblock Design principles for engaging and retaining virtual citizen
  scientists.
\newblock {\em Conservation Biology}, 30(3):562--570, 2016.

\bibitem{Whitehill2009}
Jacob Whitehill, Ting-fan Wu, Jacob Bergsma, Javier Movellan, and Paul Ruvolo.
\newblock Whose vote should count more: Optimal integration of labels from
  labelers of unknown expertise.
\newblock In Y.~Bengio, D.~Schuurmans, J.~Lafferty, C.~Williams, and
  A.~Culotta, editors, {\em Advances in Neural Information Processing Systems},
  volume~22. Curran Associates, Inc., 2009.

\bibitem{wiegmann2021opportunities}
Matti Wiegmann, Jens Kersten, Hansi Senaratne, Martin Potthast, Friederike
  Klan, and Benno Stein.
\newblock Opportunities and risks of disaster data from social media: a
  systematic review of incident information.
\newblock {\em Natural Hazards and Earth System Sciences}, 21(5):1431--1444,
  2021.

\end{thebibliography}

\end{document}